\documentclass[aps,prb,twocolumn,superscriptaddress,floatfix,showpacs]{revtex4-1}
\usepackage{graphicx}
\usepackage{dcolumn}
\usepackage{latexsym}
\usepackage{amssymb}
\usepackage{amsfonts}
\usepackage{amsmath}
\usepackage{fancybox}
\usepackage[binary,squaren]{SIunits}
\usepackage{colordvi}
\newcommand{\bg}[1]{\mbox{\boldmath$#1$\unboldmath}}

\def\be{\begin{equation}}
\def\ee{\end{equation}}
\def\ber{\begin{eqnarray}}
\def\eer{\end{eqnarray}}

\def\rv{{\bf r}}

\def\Hv{{\bf H}}
\def\vv{{\bf v}}

\def\kv{{\bf k}}
\def\qv{{\bf q}}

\def\Ev{{\bf E}}

\def\Rv{{\bf R}}
\def\Sv{{\bf S}}

\def\vv{{\bf v}}
\def\nn{\nonumber}

\begin{document}
\title{Theory of coupled spin-charge transport due to spin-orbit interaction in inhomogeneous two-dimensional electron liquids}
\author{Ka Shen}
\affiliation{Department of Physics and Astronomy, University of Missouri,
  Columbia, Missouri 65211, USA}
 \author{R. Raimondi}
\affiliation{CNISM and Dipartimento di Matematica e Fisica, Universit\`a Roma Tre,
Via della Vasca Navale 84, 00146 Rome, Italy}
\author{G. Vignale}
\affiliation{Department of Physics and Astronomy, University of Missouri,
  Columbia, Missouri 65211, USA}
  \affiliation{Donostia International Physics Center (DIPC), Manuel de Lardizabal 4, E-20018 San 
Sebasti\'an, Spain}
%\keywords{}
%72.25.Dc 	Spin polarized transport in semiconductors
%71.45.-d 	Collective effects
%75.70.Tj       Spin-orbit effects
%85.75.-d       Magnetoelectronics; spintronics: devices exploiting spin polarized transport or integrated magnetic fields
\date{\today}
\begin{abstract}
Spin-orbit interactions in two-dimensional electron liquids are responsible for many interesting transport phenomena in which particle currents are converted to spin polarizations and spin currents  and viceversa.  Prime examples are the spin Hall effect, the Edelstein effect, and their inverses.  By similar mechanisms it is also possible to partially convert an optically induced electron-hole density wave  to a spin density wave and viceversa.  In this paper we present a unified theoretical treatment of these effects based on quantum kinetic equations that include not only the intrinsic spin-orbit coupling from the band structure of the host material, but also  the  spin-orbit coupling due to an external electric field and a random impurity potential.    The drift-diffusion equations we derive in the diffusive regime are  applicable to a broad variety of experimental situations, both homogeneous and non-homogeneous, and include on equal footing ``skew scattering" and ``side-jump" from electron-impurity collisions.   As a demonstration of the strength and usefulness of the theory we apply it to the study of  several effects of current experimental interest: the inverse Edelstein effect,  the spin-current swapping effect, and the partial conversion of an electron-hole density wave to a spin density wave in a two-dimensional electron gas with  Rashba and Dresselhaus spin-orbit couplings, subject to an electric field.   
\end{abstract}
\maketitle

\section{Introduction}

During the past decade spin-orbit interactions in electron liquids have emerged  as one of the most exciting topics in  spintronics.~\cite{rmp_76_323,Fabian07,Awschalom07np,MWu09}  While classic spintronic devices (e.g., GMR read heads) rely on  strong exchange interactions between spin-polarized conduction electrons and the local magnetization of a ferromagnetic host, spin-orbit interactions offer the possibility to couple spin and charge degrees of freedom directly in a non-magnetic material.  Outstanding examples of  spin-orbital effects are the conversion of a regular electronic current into a spin current (spin Hall effect)~\cite{DPshe71,Murakami03,Sinova04,Kato04,Wunderlich05} and the generation of a non-equilibrium spin polarization  by an electronic current (Edelstein effect).~\cite{Ivchenko78,Edelstein90,Aronov89,KatoEE04,Silov04,Sih05,Silsbee04} The reciprocal effects, i.e., spin-current to current~\cite{SaitohISHE06,KAndo12} and spin polarization to current conversions,~\cite{Sanchez13,Ganichev02} have also been observed---the latter being also known as spin galvanic effect.~\cite{Ivchenko89,Ivchenko90,Ganichev02}  These are potentially useful effects, which have already been  successfully employed to generate spin-currents,~\cite{Sih06} excite and detect spin waves,~\cite{KAndo09,Kajiwara10} and  apply spin-transfer torques that can reverse the orientation of the spin polarization in  memory devices.~\cite{Liu12,LiuLee12}  More subtle effects, such as the direct coupling of spin-currents leading to ``spin-current swapping" (see section \ref{SC_swapping}) have also been predicted~\cite{Lifshits09,Sdjina12} and  await experimental verification.  In addition, recent developments in transient spin grating spectroscopy~\cite{Cameron96,Weber05,Yang2011a} have opened the way to detailed studies of the coupled dynamics of spin and charge in inhomogeneous electronic structures.  For example, the diffusion of a spin density wave~\cite{Weber05} and its drift under the action of an electric field have been studied in detail,~\cite{Yang2011a} revealing interesting many-body effects; the existence of long-lived spin-helical states in GaAs quantum wells has been confirmed.~\cite{Weber05,Walser12}  In a recent paper, building on a previous suggestion by Anderson {\it et al.},~\cite{Anderson2010}  we have proposed that an electron-hole density wave  in a GaAs quantum well can be partially converted into a spin density wave by the application of a strong electric field parallel to the wavefronts.~\cite{ShenV13b}  This and similar effects are by no means confined to conventional electron layers in GaAs: inter-metallic interfaces,  layered oxides,  monolayer materials like MoS$_2$, and ``functionalized graphene"  are all promising platforms for the observation of spin-charge conversion due to strong spin-orbit interaction.   It is therefore  important to develop a broadly applicable, easy-to-use formalism for describing the coupled evolutions of spin and charge densities and their associated currents in the presence of an external electric field.  The theoretical challenge is to provide a unified treatment of the different effects, including spin precession due to intrinsic spin-orbit coupling from the band structure of the host material,  spin relaxation, spin-orbit interaction with impurities (leading to effects such as skew scattering and side jump), and spin-orbit interaction with the external electric field.

An elegant and intuitively appealing set of spin-charge coupled drift-diffusion equations, involving charge density $N$, the spin density $\mathbf S$, the charge current $\mathbf J$, and the spin current $\mathbf J^a$, was derived in Refs.~\onlinecite{Raimondi10,GoriniPRB10,Schwab10,Raimondi_AnnPhys12} from a $SU(2)$ gauge field theoretical description of the spin-orbit coupling.  
%In this paper we extend and generalize those equations to include (i) spin-orbit coupling with impurities, which is responsible {\it inter alia} for the extrinsic spin Hall effect  (ii) an external electric field  (iii) a spin-current swapping term.  
These equations have the form
\ber
 \partial_t N &=&-\partial_i J_i\,,\label{eqs1}\\
 \partial_t S^a &=&-[\nabla_i{J}_i]^a-\delta S^a/\tau_{EY}\,,\label{conti1}\\
  J_i&=&-(v_i+D\partial_i ) N - \gamma_{ij}^a J_j^a\,,\label{eqcc}\\
  J_i^a&=&-v_iS^a-D [\nabla_i  S]^a - \gamma_{ij}^a J_j+\kappa\left(J_a^i-\delta_{ai}J_l^l\right)\,,\nn\\
\label{eqsc}
\eer
where 
\be\label{CovariantDerivative}
[\nabla_i V]^a \equiv \partial_i V^a -2\epsilon^{abc}A^b_i{V}^c
\ee
is the $SU(2)$-covariant derivative of a generic vector field $V^a$ and $\delta \Sv = \Sv-\Sv_{\rm eq}$ is the deviation of the spin density from its equilibrium value, $\Sv_{\rm eq}$  (thus the theory is applicable to ferromagnetic states).  The upper index $a$ labels components in spin space, while the lower index $i$ labels components in coordinate space.  The $SU(2)$-vector potential  $A^b_i$ describes the coupling between the $b$-th component of the spin and the $i$-th component of the orbital motion.  In the above equations $D$ is the diffusion constant ($D=v_F^2\tau/d$, where $v_F$ is the Fermi velocity and $\tau$ the current relaxation time in $d$ dimensions), $v_i$ is the $i$-th component of the macroscopic drift velocity caused by an electric field $\Ev$ ($v_i=e\tau E_i/m$), and $\tau_{\rm EY}$  is the Elliot-Yafet (EY) spin relaxation time.~\cite{Elliott54,Yafet63}
%\begin{eqnarray}
%  \partial_t S^a &=&-\partial_i{J}_i^a+2\epsilon_{abc}A^b_i{J}^c_i-S^a/\tau_{s},\label{conti1}\label{eqs1}\\
%  J_i^a&=&-(v_i+ D \partial_i)  S^a +2D\epsilon_{abc}A_i^bS^c - \gamma_{ij}^a J_j,\\
%  J_i&=&-(v_i+D\partial_i ) N - \gamma_{ij}^a J_j^a,
%\end{eqnarray}
%where $A_i^b$ corresponds to the $SU(2)$ gauge field due to spin-orbit coupling with $i$ and $b$ being indexes in real space and spin, respectively. 
$\gamma_{ij}^a$ stands for the spin Hall tensor, which connects the $J^a_i$ component of the spin current to the $J_j$ component of the charge current.  Its explicit form is   $\gamma_{ij}^a=\theta_{\rm SH}\epsilon^{ija}$ (Ref.~\onlinecite{Schwab10}) where $\theta_{\rm SH}$ known as  the ``spin Hall angle": this is a direct manifestation of the $SU(2)$ magnetic field, i.e., the covariant curl of the $SU(2)$ vector potential.   $\kappa$ is the spin-current swapping constant, derived in Appendix~\ref{APP_SCS}.  Lastly,   $\epsilon^{abc}$ is the Levi-Civita antisymmetric tensor, and a sum over repeated indices is implied throughout.

Equations~(\ref{eqs1})-(\ref{eqsc}) have a transparent physical meaning.  For example, the second equation is the generalized continuity equation for the spin density. The relaxation term $-\delta S^a/\tau_{EY}$  takes into account the Elliot-Yafet (EY) spin relaxation process  resulting from the spin-orbit interaction with impurities.  
At the same time, the spin precession that occurs between electron-impurity collisions and is responsible for the D'yakonov-Perel' (DP) spin relaxation mechanism~\cite{DPsrt71} is taken into account by the vector potential term in the $SU(2)$-covariant derivative.  
The last two equations have a similarly transparent meaning: they express the (spin) current  as  a sum of drift, diffusion, and (spin) Hall currents.  In particular, as we will show below,  it is the diffusion part of the spin current  that yields the DP spin relaxation once is inserted back into the continuity equation for the spin density.  Additional source terms, such as spin injection and spin electric fields can be added to the right hand sides of these equations.~\cite{GoriniPRB10,ShenVR14}
For example, a Zeeman field $\Hv$ coupling to the spin density enters the spin continuity equation through an additional precessional term $\propto (\Hv \times \Sv)^a$  on the right hand side of Eq.~(\ref{conti1}), and an additional spin current driving term $\sigma_s (\nabla_i H)^a$ on the right hand side of Eq.~(\ref{eqsc}), where $\sigma_s$ is the homogeneous spin-current conductivity.  At the same time, the equilibrium spin-density $\Sv_{\rm eq}$ must be reinterpreted as the quasi-equilibrium spin density in the presence of the instantaneous (frozen) field $\Hv(t)$.

The application of Eqs.~(\ref{eqs1})-(\ref{eqsc})  to homogeneous spin-orbit coupled systems has demonstrable advantages over more microscopic approaches, such as non-equilibrium Green's function theory and quantum kinetic equations.    The quantities considered here -- densities and their associated currents---are all obtained as integrals of the non-equilibrium Green's function over frequency and momenta.    While the  integrated quantities contain less information than the underlying Green's function, they are more directly connected to the experimental description of the phenomena.   
%{\bf (Raimondi: I do not think that we should put in antagonism the phenomenological equations with the microscopic approach. The former follows from the latter.  Rather, one should emphasize that the SU(2) language allows to phenomenologically motivate equations, without necessarily going through the microscopic derivation.)} such as the Green's function approach, {\bf (Raimondi: Do you mean Kubo linear response approach? Green function approach, in principle, include also the non-equilibrium fromalism which is actually used to derive the continuity equations.)}
Furthermore, there are certain features of the exact kinetics that are ``hard-wired" in the macroscopic drift-diffusion equations,  whereas in the microscopic theory they only emerge from a careful enumeration of diagrams and  delicate cancellations of seemingly different terms. 
For example the infamous ``non-analyticity puzzle", whereby the spin Hall conductivity of the Rashba model appears to drop suddenly from a finite value to zero as soon as the Rashba spin-orbit coupling is turned on, is completely demystified: the EY relaxation time---a quantity of second-order in the strength of the extrinsic spin-orbit coupling---provides the energy scale against which the Rashba spin-orbit coupling must be assessed as large or small.~\cite{Raimondi_AnnPhys12}  
%The competition between the DP and EY spin relaxation may also give rise to a non-monotonic in-temperature behavior of the spin lifetime when inelastic scattering effect are taken into account.~\cite{LiuPRB12} 
In a more recent application,  the simple addition of a spin injection term to the right-hand side of Eq.~(\ref{conti1}) has enabled us to successfully analyze the inverse Edelstein effect (also known as spin-galvanic effect), i.e., the generation of charge current from a non-equilibrium spin accumulation.~\cite{Ivchenko89,Ivchenko90,Ganichev02,Sanchez13}
 The $SU(2)$ theory is also easily applicable to spin-charge conversion phenomena that occur in {\it inhomogeneous} systems.  We have in mind, in particular, the electron-hole density waves and the spin density waves that can be generated by letting two non-collinear laser beams with different polarization interfere with each other on the surface of a semiconductor quantum well.~\cite{Cameron96,Weber05}   
% The resulting inhomogeneous structures are  known as electron-hole gratings and spin gratings, respectively,   because they act as  diffraction gratings for light.  
 Recent experiments have demonstrated that it is possible to probe in real time  (on a picosecond time scale)  not only the diffusive dynamics of these inhomogeneous structures, but also their drift under the action of an externally imposed electric field.\cite{Yang2011a,Yang2012}   The (spin) Hall transport dynamics is also in principle accessible to these experimental techniques.  Experience with homogeneous transport phenomena suggests that an extended $SU(2)$ formulation  would be a very useful theoretical tool for the description of inhomogeneous systems. This paper presents such a formulation. 

 In comparison with previous derivations of spin-charge coupled  drift-diffusion equations for  two-dimensional electronic systems,~\cite{Burkov04,Shytov04,Kleinert2007,Bryksin07,Anderson2010,LiuPRB12} the present formulation is characterized, formally, by the explicit use of the $SU(2)$ covariant derivatives, and, physically, by a careful inclusion of the spin-orbit interaction between the electrons and the impurities, as well as the external electric field.     To this end, we have carefully re-derived the kinetic equation in inhomogeneous systems, taking into account the spin-orbit coupling with the impurities and the electric field to the leading order that allows us to capture effects such as ``side-jump and ``spin-current swapping", which were not included in our previous studies of inhomogeneous density/spin dynamics.~\cite{ShenV13b}

% spin precession has been problematic in the past, leading to conflicting statement in the literature.~\cite{TsePRB06,Raimondi09,Hankiewicz_PhaseDiag_PRL08}  The problem was eventually solved in Refs.~\onlinecite{Raimondi_AnnPhys12} with the recognition of the crucial role played by the Elliot-Yafet relaxation rate - a quantity of second-order in the strength of the spin-orbit coupling.    However, the side-jump term was still missing in our previous studies of inhomogeneous density/spin dynamics, as well as in previous papers devoted to that subject.
 At last, all the relevant terms are included in the  form of a generalized  drift-diffusion equation.  It is found, quite satisfactorily, that skew scattering, side jump and intrinsic  contributions  enter the spin Hall angle on  equal terms, i.e., additively.   On the other hand,  the full spin Hall conductivity cannot be simply expressed as the sum of intrinsic and extrinsic contributions for reasons that have already been discussed in the literature~\cite{Hankiewicz_PhaseDiag_PRL08,Raimondi_AnnPhys12} and will be further explained below.
The resulting equations for spin and charge densities and their currents provide a unified theoretical framework within which one can easily treat  both homogeneous and non-homogeneous spin-charge conversion phenomena, such as the spin Hall effect, the Edelstein effect,  the spin-current swapping effect,  and the partial conversion of an electron-hole density wave into a spin density wave under the application of an electric field parallel to the wavefronts.   Throughout the paper we will emphasize the main concepts and present the final results of complex calculation.  The interested reader will find the details of the derivations in the Appendices.\\

\section{Model Hamiltonian}
The theory we are going to present applies to a class of two-dimensional model Hamiltonians of the form 
%We focus for definiteness on  the two-dimensional electron gas in the conduction band of a   (001) quantum well in a semiconductors of the zincblende structure (e.g. GaAs).    The system is described by the  Hamiltonian
\begin{equation}\label{ModelHamiltonian}
  H=H_{\kv}+H_E(\rv) +H_V(\rv)\,,
\end{equation}
where
%\be\label{HK}
%H_{\kv}=\frac{1}{2m}\sum_i\left(k_i +\sum_jA_i^j\sigma^j\right)^2 
%\ee
\be\label{HK}
H_{\kv}=\frac{k^2}{2m}+\frac{1}{m}\sum_{i,j} k_i A_i^j\sigma^j
\ee
is an  effective mass Hamiltonian for electrons of momentum $\kv$,  $\sigma^j$ are Pauli matrices for the spin and $A_i^j$ are the components of a uniform spin-dependent ($SU(2)$) vector potential,  which describes both the effective spin-orbit interaction with the crystal lattice and the spin-orbit interaction with an in-plane field $\Ev$.    In addition, we have
two terms that break the conservation of crystal momentum:   
\begin{equation}\label{HE}
H_E(\mathbf r)=e\mathbf E\cdot\mathbf r\,, 
\end{equation}
is the regular interaction with an in-plane uniform electric field $\Ev$, 
and 
\begin{equation}\label{HU}
H_V(\mathbf r)=V(\mathbf r)-\alpha'{\bg \sigma}\times\nabla_{\mathbf{r}}V(\mathbf{r})\cdot(-i\nabla_{\mathbf{r}})\,, 
\end{equation}
is the complete electron-impurity potential, of which $V(\mathbf r)$ is the spin-independent part and $\alpha'{\bg \sigma}\times\nabla_{\mathbf{r}}V(\mathbf{r})\cdot(-i\nabla_{\mathbf{r}})$ the spin-orbit coupling part (only non-magnetic impurities are considered). Here          $\alpha' \equiv\lambda_c^2/4$ is the square of the effective Compton wavelength for the material under study ($\alpha'\sim 5$ \AA$^2$ in GaAs). %{\bf(for future application, consider the possibility of redefining $\alpha'$ as a dimensionless quantity $\alpha'k_F^2$)}   
The presence of the spin-orbit term in Eq.~(\ref{HU}) is essential for the extrinsic spin Hall effect.

As a concrete example, consider the case of a (001) quantum well in a semiconductor of the zincblende structure (e.g. GaAs) with Rashba and Dresselhaus interactions and an in-plane electric field $\Ev$.  Then the non-vanishing components of the $SU(2)$ vector potential are
\ber\label{QW001}
&&A_x^y=m\lambda_1\,,~~~~~~~~~A_y^x=m\lambda_2\nn\\
&& A_x^z=m\alpha'eE_y\,, ~~~~A_y^z=-m\alpha'eE_x\,,
\eer
where $\lambda_1=\alpha+\beta$ and $\lambda_2=\beta-\alpha$ with $\alpha$ and $\beta$ being the Rashba~\cite{Rashba84} and Dresselhaus~\cite{Dresselhaus55} SOC coefficients separately.  Following common usage, the  $x$ and $y$ axes are defined in the [110] and [$\bar 1$10] directions respectively. The two terms on the last line describe the spin-orbit interaction with the in-plane electric field.

%Taking into account the band SOCs due to the Rashba~\cite{Rashba84} and linear Dresselhaus~\cite{Dresselhaus55} SOCs, and the external electric field ${\bf E}$, we write the ideal crystal Hamiltonian for the electron with momentum $\mathbf k$ as
%\begin{equation}
%H_{\mathbf{k}} = \varepsilon_k+k_{x}\lambda_{1}\sigma_{y}+k_{y}\lambda_{2}\sigma_{x}+e\mathbf{E}\cdot(\alpha'{\bg \sigma}\times\mathbf{k}), \label{HK}
%\end{equation}
%where $\lambda_1=\alpha+\beta$ and $\lambda_2=\beta-\alpha$ with $\alpha$ and $\beta$ being the Rashba and Dresselhaus SOC coefficients separately, and $\varepsilon_k={k^{2}}/{(2m)}$ stands for the kinetic energy ($m$ is the effective mass). Note that the $x$ and $y$ axes are defined in the [110] and [$\bar 1$10] directions throughout this paper. The last term in Eq.~(\ref{HK}) represents the SOC induced by the external electric potential. $\alpha' \equiv\lambda_c^2/4$ is the square of the effective Compton wavelength for the semiconductor ($\sim 5$ \AA$^2$ in GaAs). 

For future use, we also define the crystal and electric-field-induced spin-orbit interaction Hamiltonian $H^{\rm soc}_{\kv}$ as follows:
\be
H^{\rm soc}_{\kv} \equiv \frac{1}{m}\sum_{ij} k_iA_i^j\sigma^j\,,
\ee 
such that
\be
H_{\kv}=\varepsilon_k+H^{\rm soc}_{\kv}\,,
\ee
where $\varepsilon_k = \frac{k^2}{2m}$. 
%Ka: use this if we choose Eq.(8).}
%\be
%\varepsilon_k = \frac{k^2}{2m}+\sum_{ij}\frac{(A_i^j)^2}{2m}\,.
%\ee
%Notice that the last term on the right hand side is  a mere constant, which can and will be safely  ignored in the following discussions.

%where, in the $SU(2)$ frame, $H_{\mathbf k}^{\rm soc}=(k_i/m)A_i^j\sigma^j$ with $A_x^y=m\lambda_1$, $A_y^x=m\lambda_2$, ${A}_x^z=m\alpha'eE_y$, and $A_y^z=-m\alpha'eE_x$.
%The last two terms in Eq.~(\ref{ModelHamiltonian}) depend on $\mathbf r$ and thus break the conservation of crystal momentum.  $H_E$ represents the standard coupling with the electric field,
%\begin{equation}\label{HE}
%H_E(\mathbf r)=e\mathbf E\cdot\mathbf r\,, 
%\end{equation}
%and $H_U$ is the complete impurity potential 
%\begin{equation}\label{HU}
%H_U(\mathbf r)=V(\mathbf r)-\alpha'{\bg \sigma}\times\nabla_{\mathbf{r}}V(\mathbf{r})\cdot(-i\nabla_{\mathbf{r}})\,, 
%\end{equation}
%of which $V(\mathbf r)$ is the spin-independent part and $\alpha'{\bg \sigma}\times\nabla_{\mathbf{r}}V(\mathbf{r})\cdot(-i\nabla_{\mathbf{r}})$ the SOC part.  The presence of SOC with the impurities is essential for the extrinsic spin Hall effect.

\section{Kinetic equation} \label{SecKE}
Our starting point is the well-known~\cite{Shytov04,Rammer_qkt,MWu09} kinetic equation for the quasi-classical (Wigner) distribution function $\rho_{\mathbf k}(\rv,t)$:
\begin{equation} \label{KERho}
\partial_{t}\rho_{\kv}+i[H_{\kv},\rho_{\kv}]+\frac{1}{2}\{\nabla_{\kv}H_{\kv},\nabla_{\rv}\rho_{\kv}\}-e\mathbf{E}\cdot\nabla_{\kv}\rho_{\mathbf{k}}=I_{\kv}\,.
\end{equation}
All the quantities in this equation, including the collision integral $I_{\kv}$,  are functions of a position $\rv$ and a  time $t$, which are not explicitly written down.
Here the symbols $[\ , \ ]$ and $\{\ , \}$ stand for commutator and anticommutator, respectively and all quantities are matrices in spin space.
The collision integral $I_{\mathbf k}$ arises from the interaction with impurities, Eq.~(\ref{HU}), and is expressed in terms of the contour-ordered Green's function $G_{\kv}(\rv,t,t')$ and the self-energy $\Sigma_{\kv}(\rv,t,t')$  as follows
\begin{equation}
I_{\mathbf k}(t)=-\left(\int_c dt^\prime[\Sigma_{\mathbf{k}}(t,t^\prime)G_{\mathbf{k}}(t^\prime,t)-G_{\mathbf{k}}(t,t^\prime)\Sigma_{\mathbf{k}}(t^\prime,t)]\right)^{<}\label{collisionG}
\end{equation}
where the superscript $<$ denotes the lesser component of the contour integral.  Details of the derivation can be found in Refs.~\onlinecite{haug_qkt} and \onlinecite{ShenTW11}. 

%Following the standard procedures in non-equilibrium Green function approach,~\cite{haug_qkt,ShenTW11} the kinetic equation of Keldysh Green function, up to first-order in the gradient expansion, reads~\footnote{The first order gradient expansions for the collision terms, in the form of $\frac{i}{2}\int_{c}dt_{3}\{\nabla_{\mathbf{r}}\Sigma(\mathbf{r},\mathbf{k}),\nabla_{\mathbf{k}}G(\mathbf{r},\mathbf{k})\}$ and $\frac{i}{2}\int_{c}dt_{3}\{\nabla_{\mathbf{k}}\Sigma(\mathbf{r},\mathbf{k}),\nabla_{\mathbf{r}}G(\mathbf{r},\mathbf{k})\}$, have been discarded because they both vanish in the kinetic equation under generalized Kadanoff-Baym ansatz.}
%\begin{eqnarray}
%i\partial_{T}G_{\mathbf k}^<  &=&  [H_{0}(\mathbf{k}),G_{\mathbf k}^<]-\frac{i}{2}\{\mathbf{v}_{\mathbf{k}},\nabla_{\mathbf{r}}G_{\mathbf k}^<\}\nonumber \\
%&&\hspace{-0.2cm}\mbox{}+ie\mathbf{E}\cdot\nabla_{\mathbf{k}}G_{\mathbf k}^<
%  +\int_{c}dt_{3}[\Sigma_{\mathbf k}G_{\mathbf k}-G_{\mathbf k}\Sigma_{\mathbf k}]^<,
%\label{KEG}
%\end{eqnarray}
%where $G_{\mathbf k}^<(\mathbf r)$ is the local lesser Green function at $\mathbf r$ in real space, with the momentum $\mathbf k$ and $\Sigma_{\mathbf k}(\mathbf r)$ is the corresponding self-energy. Here, all the time variables in Green function and self-energy are omitted for short. Details can be found in Refs.~\onlinecite{haug_qkt,ShenTW11}. 
%{\bf Define notation, variable and contour of integration}

\begin{figure}
  \includegraphics[width=6cm]{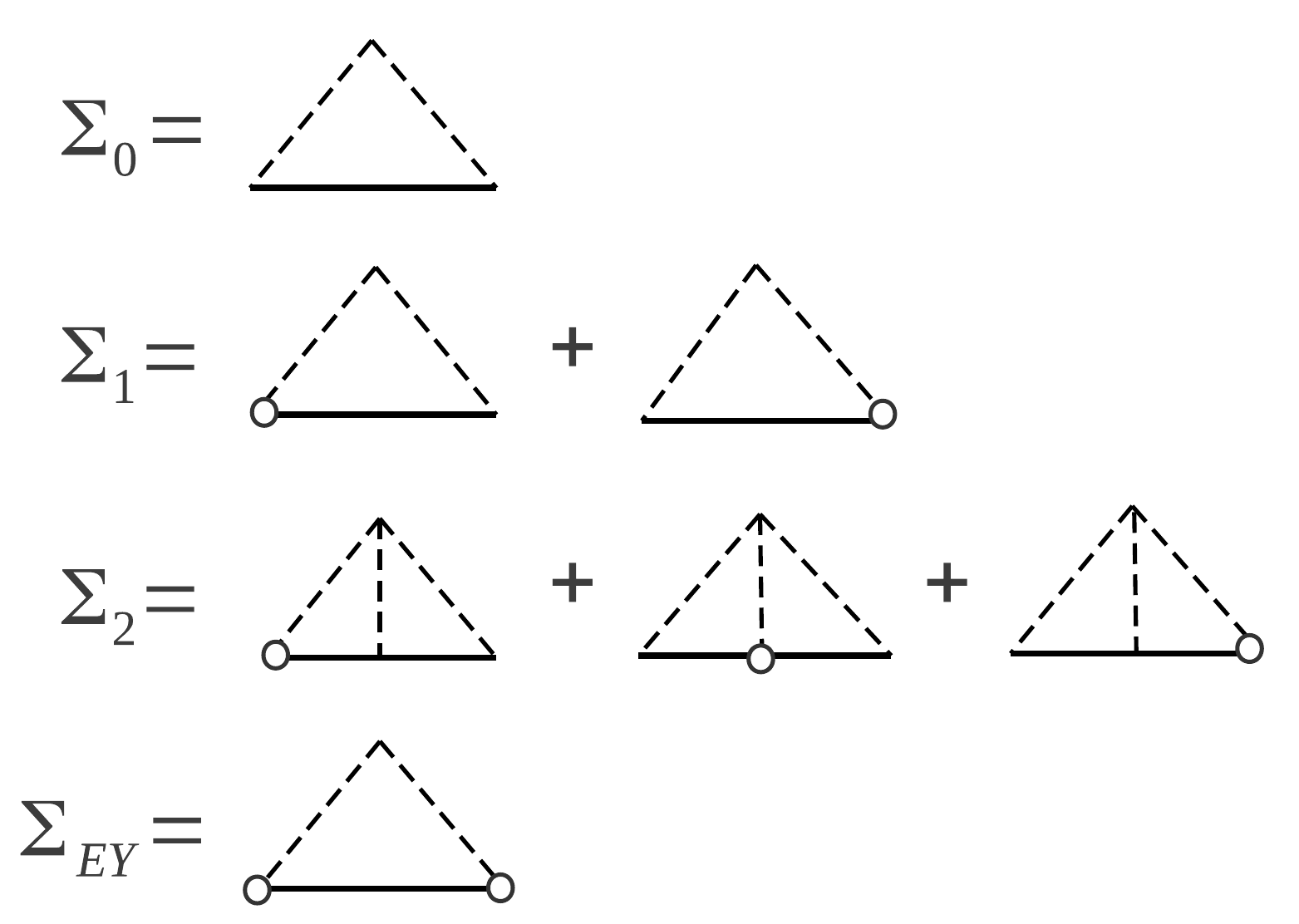}
\caption{(Color online) Diagrams for the impurity-averaged self-energy $\Sigma_0$, $\Sigma_1$, $\Sigma_2$, and $\Sigma_{\rm EY}$. The dashed line denotes the impurity averge and
%due to impurity potential, where 
the vertices with open circles represent the impurity-induced spin-orbit coupling. }
\label{figSE}
\end{figure}

The self-energy due to the impurity potential consists of four terms, which are graphically represented in Fig.~\ref{figSE}.  For a short-range $\delta$-correlated disorder potential $V(\rv)=\sum_i v_0\delta(\rv-\Rv_i)$, where $\Rv_i$ are the random position of  impurities with average density $n_i$,   these diagrams have the following analytic expressions (see Ref.~\onlinecite{Raimondi10}):
\be\label{Sigma0}
\Sigma_{0\kv}=n_{i}v_0^2\sum_{\kv'}G_{\kv'}\,,
\ee
(i.e., the usual Born approximation)
\ber\label{Sigma1}
\Sigma_{1\kv}
&=& n_{i}v_0^2\alpha'\sum_{\mathbf{k}'}(-i[{\bg \sigma}\cdot\mathbf{k}\times\mathbf{k}',G_{\mathbf k'}]\nonumber\\
&&\hspace{-0.2cm}\mbox{}-\frac{1}{2}\{{\bg\sigma}\times(\mathbf{k}-\mathbf{k}'),\nabla_{\rv}G_{\mathbf k'}\})\,,
\eer
and
\ber\label{Sigma2}
\Sigma_{2\mathbf k}
& = & -in_{i}v_0^3\alpha'\sum_{\mathbf{k}',\mathbf k''}(\mathbf k\times \mathbf k'\cdot {\boldsymbol\sigma} G_{\mathbf k'}G_{\mathbf k''}\nonumber\\
&&\hspace{-0.2cm}\mbox{}+G_{\mathbf k'}\mathbf k'\times \mathbf k''\cdot{\boldsymbol\sigma} G_{\mathbf k''}+G_{\mathbf k'}G_{\mathbf k''}\mathbf k''\times \mathbf k\cdot{\boldsymbol\sigma})\,.\nn\\  \label{sigma2}
\eer
The gradient term in $\Sigma_{1\kv}$ comes from the derivative operator in Eq.~(\ref{HU}) acting on the spatial argument of the Green function.  A similar gradient term in $\Sigma_{2\kv}$ is neglected on account of its smallness in a system with smooth inhomogeneity.
% There is no gradient term in $\Sigma_2$ to the order of accuracy at which we are working.
The last diagram, denoted by $\Sigma_{\rm EY}$ is given by~\cite{Raimondi10}
\be\label{SiñgmaEY}
\Sigma_{\rm EY\kv}=n_i v_0^2 (\alpha')^2\sum_{\mathbf k'}\sigma^z G_{\mathbf k'}\sigma^z(\mathbf k\times\mathbf k')_z^2
\ee
and is responsible for Elliot-Yafet spin relaxation.

%for short-range disorder, two additional terms of first order in the coupling constant $\alpha'$, which we denote by $\Sigma_{1\mathbf k}$ and $\Sigma_{2\mathbf k}$ respectively.  $\Sigma_{1\mathbf k}$ is still of second order in $V$ and contributes to the ``side-jump" effect.  $\Sigma_{2\mathbf k}$ is of third order in $V$ and is responsible for the skew-scattering effect.  The diagrams for these contributions are shown in Ref.~\onlinecite{Raimondi10}, where the expressions read

To write out the collision integral in terms of the density matrix $\rho_{\kv}$, we employ the standard rules of analytic continuation~\cite{haug_qkt} combined with the generalized Kadanoff-Baym ansatz,~\cite{GKB86,haug_qkt} which expresses the lesser or greater components of the Green's function (and hence the self-energy) in terms of equal-time Green's functions and retarded/advanced propagators:
\begin{eqnarray}
G_{\mathbf k}^{\gtrless }(\mathbf r,t,t_{1})&=&\mp i[\theta(t-t_{1})e^{-i\tilde H(t-t_{1})}\rho_{\mathbf k}^{\gtrless }(\mathbf r, t_{1})\nonumber\\
  &&+\theta(t_{1}-t)\rho_{\mathbf k}^{\gtrless }(\mathbf r,t)e^{-i\tilde H(t-t_{1})}]\,.\label{GKB}
\end{eqnarray}
Here $\rho_{\mathbf k}^<=\rho_{\mathbf k}$ and $\rho_{\mathbf k}^>=1-\rho_{\mathbf k}$.  Notice that the Hamiltonian $\tilde H$ contains not only $H_{\mathbf k}$, but also the electric potential $H_E$, i.e.,  $\tilde H(\mathbf r,\mathbf k)=H_{\mathbf k}+H_E(\mathbf r)$. According to Refs.~\onlinecite{Culcer10} and \onlinecite{Bi13}, the inclusion of $H_E$ is essential to capture the full side-jump effect.
% for which we look at the collision integral in the second order in $v_0$, corresponding to self energy from $\Sigma_0$ and $\Sigma_1$.

%By substituting the Hamiltonian and self-energy into Eq.~(\ref{KEG}), we finally reach the kinetic equation of the density matrix
%\begin{equation}
%\partial_{t}\rho_{\mathbf{k}}+i[H_{\mathbf{k}},\rho_{\mathbf{k}}]+\frac{1}{2}\{\mathbf{v}_{\mathbf{k}},\nabla_{\mathbf{r}}\rho_{\mathbf{k}}\}-e\mathbf{E}\cdot\nabla_{\mathbf{k}}\rho_{\mathbf{k}}=I_{\mathbf k}, \label{KERho}
%\end{equation}
%{\bf [what is $H_{\bf k}$?]}  where $I_{\mathbf k}$ on the right-hand side is the collision integral
%\begin{equation}
%I_{\mathbf k}(t)=-\int_c dt_{1}[\Sigma_{\mathbf{k}}(t,t_{1})G_{\mathbf{k}}(t_{1},t)-G_{\mathbf{k}}(t,t_{1})\Sigma_{\mathbf{k}}(t_{1},t)]^{<}.
%\end{equation}
Treating the SOCs and electric potential as small perturbations, we expand the propagator in Eq.~(\ref{GKB}) as
\begin{equation}\label{EXP}
  e^{-i\tilde H\tilde t}\approx e^{-i\varepsilon_{k}\tilde t}(1-i\tilde t H_{\mathbf k}^{\rm soc}-i\tilde t H_E)\,,
\end{equation}
where $\tilde t \equiv t-t_1$.
To construct the collision integral we separately consider the contributions arising from the first, the second, and the third term in the  brackets in Eq.~(\ref{EXP}).

\subsection{Collision integral from unperturbed propagator} 
The first term in the brackets in Eq.~(\ref{EXP}), when substituted in Eq.~(\ref{GKB}) and subsequently in the expression~(\ref{collisionG}) for the collision integral leads to 
\ber
\hspace{-0.3cm}I_{\mathbf{k}}^{(0)} & =& n_{i}v_0^2\sum_{\mathbf{k}^{\prime}}2\pi\delta(\varepsilon_{k}-\varepsilon_{{k}^{\prime}})\big[(\rho_{\mathbf{k}^{\prime}}-\rho_{\mathbf{k}})\nonumber\\
&&\hspace{-0.8cm}\mbox{}
-i\alpha'[{\bg\sigma}\cdot\mathbf{k}\times\mathbf{k}^{\prime},\rho_{\mathbf{k}^{\prime}}]-\tfrac{1}{2}\{\alpha'\bg\sigma\times(\mathbf{k}-\mathbf{k}'),\partial_{\mathbf{r}}\rho_{\mathbf{k}'}\}
\big]\nn\\
\eer
 (from $\Sigma_0$ and $\Sigma_1$).
Here the first term, proportional to $(\rho_{\mathbf k'}-\rho_{\mathbf k})$, arises from $\Sigma_{0\kv}$ and describes ordinary electron-impurity scattering processes. The remaining two terms arise from the two parts of  $\Sigma_{1\kv}$ in Eq.~(\ref{Sigma1}) and are  less familiar.  The second term describes the spin precession that occurs during a single electron-impurity collision, with the precession angle depending on the relative angle between the momenta before and after collision.  This spin precession, as we will show later, can manifest itself through the  ``spin current swapping".~\cite{Lifshits09} The third term with the gradient of the density matrix gives rise to a spin-density coupling proportional to the change in momentum. 
%By employing 
By introducing the standard relaxation time
%approximation
, i.e.,  
\be\label{RelaxationTime}
n_{i}v_0^2\sum_{\mathbf{k}^{\prime}}2\pi\delta(\varepsilon_{k}-\varepsilon_{{k}^{\prime}})=n_{i}mv_0^2=\frac{1}{\tau}\,,
\ee
($\hbar=1$) we obtain 
\begin{eqnarray}\label{IK0}
I_{\kv}^{(0)} & =& -\frac{\rho_{\mathbf{k}}-\rho_{{k}}}{\tau}-\frac{1}{\tau}\int \frac{d\theta_{\mathbf k'}}{2\pi}i\alpha'[{\bg\sigma}\cdot\mathbf{k}\times\mathbf{k}^{\prime},\rho_{\mathbf{k}^{\prime}}]\nonumber\\
&&\hspace{-0.2cm}\mbox{}
+\tfrac{1}{2}\{\alpha'\bg\sigma\times(\mathbf{k}-\mathbf{k}'),\partial_{\mathbf{r}}\rho_{\mathbf{k}'}\},
\end{eqnarray}
where 
%$\tau$ is the standard momentum relaxation time and 
$\rho_k$ corresponds to the angular average of  $\rho_{\kv}$ over wave vectors of fixed magnitude $|\kv|=k$.
%which is suitable for our perturbation approach below.
Similarly, the  self-energy $\Sigma_2$ %{, \bf combined with the first term in the propagator~(\ref{EXP})}   
generates the skew-scattering contribution  $I^{\rm ss}_{\kv}$ to the collision integral~\cite{ChengJPCM,Raimondi_AnnPhys12}
%.  {\bf This is derived in the Appendix} and displayed here
\begin{equation}\label{ISS}
  I^{\rm ss}_{\kv}=-n_i\alpha'\frac{m^2 v_0^3}{2}\int \frac{d\theta_{\mathbf k'}}{2\pi}\{\mathbf k\times\mathbf k'\cdot{\bg \sigma},\rho_{\mathbf k'} \}\,,
\end{equation}
%Lastly 
and the self-energy $\Sigma_{\rm EY}$ generates the EY spin-relaxation contribution to the collision integral:
\be\label{IEY}
I^{\rm EY}_{\kv}=-\frac{1}{\tau}(\alpha')^{2}\int\frac{d\theta_{\mathbf{k}'}}{2\pi}(\mathbf{k}\times\mathbf{k}')_{z}^{2}(\rho_{\mathbf{k}}-\sigma^{z}\rho_{\mathbf{k}'}\sigma^{z}).
\ee

\subsection{Collision integral from first-order SOC correction to the propagator}
The second term in the brackets in Eq.~(\ref{EXP}), when substituted in Eq.~(\ref{GKB}) and subsequently in the expression~(\ref{collisionG}) for the collision integral generates  two terms related to $\Sigma_{0\kv}$ and $\Sigma_{1\kv}$, which we denote by $I_{\kv}^{(a)}$ and $I_{\kv}^{(b)}$ respectively.  Their analytic expressions are  
%To elucidate the effect of the SOC and electric potential within a single collision process, we calculate the correction due to the second and third terms in Eq.~(\ref{EXP}). According to the different order in $\alpha'$, we separate the correction associated $H^{\rm soc}$ into two parts,
\begin{equation}
I_{\kv}^{(a)} = n_{i}v_0^2\sum_{\mathbf{k}^{\prime}}\pi\{H^{\rm soc}_{\mathbf{k}}-H^{\rm soc}_{\mathbf{k}^{\prime}},\rho_{\mathbf{k}^{\prime}}-\rho_{\mathbf{k}}\}\partial_{\varepsilon_{k}}\delta(\varepsilon_{k}-\varepsilon_{{k}^{\prime}}) 
\label{IA}
\end{equation}
(from $\Sigma_0$) and
\begin{eqnarray}\label{IB}
I_{\kv}^{(b)} & = & n_{i}v_0^2\sum_{\mathbf{k}^{\prime}}\pi\big[\{H_{\mathbf{k}}^{\rm soc},[(-i\alpha'\vec{\sigma}\cdot\mathbf{k}\times\mathbf{k}^{\prime}),\rho_{\mathbf{k}^{\prime}}]\}\nonumber\\
 &  & \mbox{}+\{\rho_{\mathbf{k}},[(-i\alpha'\vec{\sigma}\cdot\mathbf{k}\times\mathbf{k}^{\prime}),H_{\mathbf{k}^{\prime}}^{\rm soc}]\}\nonumber\\
 &  & \mbox{}+[i\alpha'\vec{\sigma}\cdot\mathbf{k}\times\mathbf{k}^{\prime},\{\rho_{\mathbf{k}^{\prime}},H_{\mathbf{k}^{\prime}}^{\rm soc}\}]\big]\partial_{\varepsilon_{k}}\delta(\varepsilon_{{k}}-\varepsilon_{{k}^{\prime}})\,,\nn\\
\end{eqnarray}
(from $\Sigma_1$).  We notice that $I^{(a)}_{\mathbf k}$  simply describes the SOC-induced shift in the single particle energies that enter the $\delta$-function of conservation of energy.  
In the relaxation time approximation, 
%{\ROB After integration by parts and exploiting the fact that $H^{\rm soc}$ is an odd function of $\kv$,}
this term can be readily evaluated as~\cite{ShenV13b}
\begin{equation}\label{IA2}
I_{\kv}^{(a)} = \frac{1}{2\tau}\{H^{\rm soc}_{\kv},\partial_{\varepsilon_{k}}\rho_{k}\}.
\end{equation}
%which has the standard form of the relaxation time approximation. 
In contrast to this, 
%the relaxation time approximation 
a similar relaxation time approximation form cannot be 
%applied to 
derived for $I_{\kv}^{(b)}$ in a simple way, because of the complicated dependence on both $\mathbf k$ and $\mathbf k'$. Fortunately, this part is already of the  order that is required for the description of the  side-jump effect, namely first order in band SOC together with first order in $\alpha'$, which allows us to extract its contribution to the drift-diffusion equation,  as detailed in Appendix~\ref{APP_DDE}. One may notice that the second and third terms in $I^{(b)}_{\kv}$ describe  the ``anomalous spin precession'' introduced in Ref.~\onlinecite{Bi13}, here generalized to spin-polarized distributions.

\subsection{Collision integral from electric field correction to the propagator}
Finally, the electric potential correction to the propagator, i.e. the third term in the brackets of Eq.~(\ref{EXP}), when combined with the self-energy $\Sigma_{1\kv}$, leads to a contribution to the collision integral of the form
\ber\label{IC}
I_{\mathbf{k}}^{(c)} & = & n_{i}v_0^2\sum_{\kv^{\prime}}\pi\{\alpha'e\mathbf{E}\cdot{\bg \sigma}\times(\mathbf{k}-\kv^\prime),\rho_{\kv^{\prime}}-\rho_{\kv}\}\nonumber\\
&&\hspace{-0.2cm}\mbox{}\times[\partial_{\varepsilon_{k}}\delta(\varepsilon_{k}-\varepsilon_{k^{\prime}})]\,.
\eer
By comparison with Eq.~(\ref{IA}) one can see that this term equals the contribution due to electric-field-induced SOC in $I_{\mathbf k}^{(a)}$, reflecting the famous factor of ``2'' in the side jump effect.~\cite{Engel05,Tse06,Culcer10} 
In the relaxation time approximation 
%{\ROB After integration by parts} 
it takes the form
\begin{equation}\label{IC2}
I_{\mathbf{k}}^{(c)} = \frac{1}{2\tau}\{\alpha'e\mathbf{E}\cdot{\bg \sigma}\times\mathbf{k},\partial_{\varepsilon_{k}}\rho_{k}\}\,.
\end{equation}
The remaining parts of the self-energy,  $\Sigma_2$ and the gradient term in $\Sigma_1$, combined with the second and third terms in the brackets in Eq.~(\ref{EXP}),  give  higher order contributions, which are therefore neglected.
%with $H^{\rm soc,E}_{\mathbf{k}}=\alpha'e\mathbf{E}\cdot{\bg \sigma}\times\mathbf{k}$.

\subsection{Full collision integral}
Our complete expression for the collision integral is therefore
\be\label{CollisionIntegralComplete}
I_{\kv}=I^{(0)}_{\kv} +I^{(a)}_{\kv}+I^{(b)}_{\kv}+I^{(c)}_{\kv}+I^{\rm ss}_{\kv}+I^{\rm EY}_{\kv}\,.
\ee
Compared to our previous calculation in Ref.~\onlinecite{ShenV13b}, where the collision term was given by 
\begin{equation}\label{CollisionIntegralOld}
I_{\kv}=-\frac{\rho_{\mathbf{k}}-\rho_{{k}}}{\tau}+\frac{1}{2\tau}\{H^{\rm soc}_{\mathbf{k}},\partial_{\varepsilon_{\mathbf k}}\rho_{k}\}+I^{\rm ss}_{\kv},
\end{equation}
our careful treatment with impurity-induced SOC has produced several additional contributions, such as $I^{(b)}_{\kv}$, $I^{(c)}_{\kv}$, and $I^{\rm EY}_{\kv}$, as well as the second and the third terms in $I^{(0)}_{\kv}$.  Such terms describe  interesting physical effects, e.g.,  the spin precession within a single collision due to SOC induced by impurity potential and/or band SOC. Even though some of these terms had been obtained and discussed separately for specific systems in literature,~\cite{Lifshits09,Raimondi_AnnPhys12,Bi13} our derivation pulls all the pieces together while  supplying a more general and complete kinetic theory for spatially inhomogeneous system. 

\section{Drift-diffusion equations}
The density matrix provides a  microscopic description of transport---one in which we  keep track of the detailed distribution of electrons in momentum space. Such a detailed description is often unnecessary.  For example, when studying spintronic devices we are usually interested in equations that connect  the spin and charge currents to the corresponding densities in space: information about the momentum distribution of the particles is discarded.  We are thus facing the task of reducing the kinetic equations to more manageable equations for the charge and spin densities and current densities.  Such equations are referred to as ``drift-diffusion equations", and this section presents the main steps in their derivation.   More precisely, in subsection~\ref{EDensity}, we derive the equations that govern the evolution of the densities, without explicit reference to the currents.    Explicit formulas for the currents are derived in subsection~\ref{ECurrent}.

\subsection{Equations for the densities} \label{EDensity}
We begin by expanding the density matrix as $\rho_{\kv}=g_{\kv}^i\sigma^i$ and $\rho_{k}=g_{k}^i\sigma^i$.  Here, in addition to the familiar Pauli matrices $\sigma^1 = \sigma_x$,  $\sigma^2 = \sigma_y$ and $\sigma^3 = \sigma_z$, we have also included, for convenience,  the $2\times 2$ identity matrix $\sigma^0 = {\bf 1}$.  Thus $g_{\kv}^0$ represents the charge distribution regardless of spin orientation. Similarly, we expand the collision integral as $I_{\kv}= I_{\kv}^i\sigma^i$.

After Fourier transformation with respect to $t$ and $\rv$, with conjugate variables $\omega$ and $\qv$ respectively,  the kinetic equation is rewritten as
\begin{equation}\label{KE2}
({\cal I}+{\cal K}_{{\bf k}})
  {\bf g}_{{\bf k}}
 = ({\cal I}+{\cal T}_{{\bf k}})
{\bf g}_{k}+\int \frac{d\theta_{\mathbf k'}}{2\pi}{\cal M}_{\mathbf{k},\mathbf{k}'}
{\bf g}_{{\bf k'}} +\tau {\bf I}_{\mathbf{k}}^{(b)}\,, 
\end{equation}
where ${\bf g}$ is a column vector with components $(g^0,g^1,g^2,g^3)$ and ${\bf I}$ is also a column vector with components $(I^0,I^1,I^2,I^3)$.
Here, ${\cal I}$ is the $4\times 4$ identity matrix and ${\cal K}_{\kv}$, defined in Appendix~\ref{APP_Matrix}, generates what is essentially  the scattering-free dynamics of the density matrix. The right-hand side of Eq.~(\ref{KE2}) includes all the relevant collision terms derived in the previous section, with the matrices ${\cal T}_\kv$ and ${\cal M}_{\kv,\kv'}$ defined in Appendix~\ref{APP_Matrix}.  Notice that, due to the spatial Fourier transformation, the matrices  ${\cal K}_{\mathbf k}$, ${\cal T}_\kv$ and ${\cal M}_{\kv,\kv'}$, which were previously functions of $\rv$, have now become functions of the conjugate wavevector $\qv$ (see Appendix~\ref{APP_Matrix}).

 In the limit in which the extrinsic SOC constant $\alpha'$  vanishes the kinetic equation can be solved (by exploiting its 
 %in the 
 relaxation time approximation form) yielding 
\be\label{QuasiEquilibrium}
{\bf g}_{\kv} (\alpha'=0)=[({\cal I}+{\cal K}_{\bf k})^{-1}({\cal I}+{\cal T}_{\mathbf k})]{\bf g}_k \equiv \tilde {\bf g}_{\kv}.
\ee
Notice that the full angle dependence (direction of $\kv$) of ${\bf g}_{\kv}$ is entirely determined by the $\kv$-dependence of the matrices ${\cal K}_{\bf k}$ and
${\cal T}_{\mathbf k}$. Hence, in this limit of vanishing extrinsic SOC, by taking the angle average over $\kv$ in Eq.(\ref{QuasiEquilibrium}) one can obtain a closed equation for the angle averaged density matrix vector ${\bf g}_k$.
In the diffusive regime the relaxation time is very short compared to the  time scale over which the distribution function varies significantly. Therefore the effect of $\alpha'$ is a small correction to the collision integral.  Substituting ${\bf g}_{\kv} \simeq \tilde {\bf g}_{\kv}$  into the scattering terms on the right hand side of  Eq.~(\ref{KE2}) yields 
%\begin{eqnarray}\label{gk_tot}
%  {\bf g}_{\mathbf k}&\simeq & ({\cal
%      I}+{\cal K}_{\bf k})^{-1}[({\cal I+T}_{\bf
%    k}){\bf g}_k
%  + \int \frac{d\theta_{\mathbf k'}}{2\pi}{\cal
%      M}_{\mathbf k,\mathbf k'}{\tilde {\bf g}}_{\mathbf k'}]
% \nonumber\\
% &&+\tau ({\cal I}+{\cal K}_{\bf k})^{-1}I_{\mathbf{k}}^{(b)}({\tilde {\bf g}}_{\mathbf k},{\tilde {\bf g}}_{\mathbf k^\prime})\,.
%\end{eqnarray}
\be\label{gk_tot}
  {\bf g}_{\mathbf k}\simeq  {\tilde {\bf g}}_{\mathbf k}+({\cal
      I}+{\cal K}_{\bf k})^{-1}\left[\int \frac{d\theta_{\mathbf k'}}{2\pi}{\cal
      M}_{\mathbf k,\mathbf k'}{\tilde {\bf g}}_{\mathbf k'}+\tau I_{\mathbf{k}}^{(b)}({\tilde {\bf g}}_{\mathbf k},{\tilde {\bf g}}_{\mathbf k^\prime})\right].
\ee
Now all terms on the right hand side of the above equation are linear functions of the angle averaged density matrix vector ${\bf g}_k$, the coefficient depending on $\kv$ after integration over $\kv'$.
%\begin{eqnarray}
%  g_{\mathbf k}^{i}&\simeq &\langle [{\cal I}+({\cal I}+{\cal K}_{\bf k'})^{-1}{\cal
%      M}_{\mathbf k',\mathbf k}] 
%  ({\cal
%    I}+{\cal K}_{\bf k})^{-1}({\cal I+T}_{\bf
%    k})\rangle^{ij}g_k^{j}\nonumber\\
%  &&+\tau \langle[({\cal I}+{\cal K}_{\bf k})^{-1}]^{ij}I_{\mathbf{k}}^{b,j}({\tilde g}_{\mathbf k},{\tilde g}_{\mathbf k^\prime})\rangle. 
%\end{eqnarray}
From this we  derive closed equations of motion for the charge ($ N=\sum_{\kv} g^0_k$)
and spin ($S^i=\sum_{\kv} g^i_k$ with $i=1$, $2$, $3$) densities by doing the appropriate sums over $\kv$.  
In the diffusive regime,
%limit,  
$\omega\tau\ll 1$, we do a linear expansion with respect to $\omega$ and transform back from $\omega$  to $t$, to get the diffusion equation
%\begin{equation}\label{DDE}
%\partial_t (\Delta N,S_x,S_y,S_z)^T=-{\cal D}(\tilde {\mathbf q})(\Delta N,S_x,S_y,S_z)^T\,.
%\end{equation}
\begin{equation}\label{DDE}
\partial_t \left(\begin{array}{c}
\Delta N_\qv\\
S^x_\qv\\
S^y_\qv\\
S^z_\qv\end{array}\right)
=-{\cal D}(\qv)\left(\begin{array}{c}
\Delta N_\qv\\
S^x_\qv\\
S^y_\qv\\
S^z_\qv\end{array}\right)\,,
\end{equation}
where $\Delta N_\qv$ and $S^i_\qv$ are, respectively,   the components of the density and the spin density deviations from the (uniform) equilibrium state with wave vector $\mathbf q$. 
\begin{widetext}The $4\times 4$  diffusion matrix $\cal D(\qv)$ is defined by
\be\label{Dq}
{\cal D}_{ij}=\frac{\delta_{ij}}{\tau}-\frac{1}{\tau}\left\langle [{\cal I}+({\cal I}+{\cal K}_{\bf k'})^{-1}{\cal
      M}_{\mathbf k',\mathbf k}]  ({\cal I}+{\cal K}_{\bf k})^{-1} ({\cal I+T}_{\bf
    k})\right\rangle_{ij}-\left.\frac{1}{S_{\mathbf q}^j}\left\langle [({\cal I}+{\cal K}_{\bf k})^{-1}]^{il} I_{\mathbf{k}}^{(b),l}({\tilde {\bf g}}_{j\mathbf k},{\tilde {\bf g}}_{j\mathbf k^\prime})\right\rangle\right\vert_{\omega=0}\,.
\ee
where $\langle. \rangle$ represents the average over the carrier distribution in momentum space and includes integration over both $\kv$ and $\kv'$.  In the last term, ${\tilde {\bf g}}_{j\mathbf k}$ denotes the contribution to the quasi-equilibrium distribution arising, in accordance with  Eq.~(\ref{QuasiEquilibrium}),  from the $j$-th component of the equilibrium density matrix ${\bf g}_k$,  i.e., ${\tilde { g}}^i_{j\mathbf k}=[({\cal I}+{\cal K})^{-1}({\cal I}+{\cal T}_{\mathbf k})]^{ij}g_k^j$. Thus, the diffusion matrix is independent of $\kv$ and $\kv'$, but it does depend, via the matrices  ${\cal K}_{\mathbf k}$, ${\cal T}_\kv$ and ${\cal M}_{\kv,\kv'}$ (see Appendix~\ref{APP_Matrix})  on the wave vector $\qv$, conjugate to $\rv$.  %Here, $S_{\mathbf q}^0$ is defined  as the amplitude of the density grating, i.e., $\Delta N_{\mathbf q}$.
In order to simplify the expression, we have introduced $S_{\mathbf q}^0=\Delta N_{\mathbf q}$ in Eq.~(\ref{Dq}) for $j=0$.
\\

In the following we assume $k_F q/m,|\lambda_i|k_F\ll E_F$ and do a perturbation expansion with respect to ${\cal T}_{\mathbf k}$ and ${\cal K}_{\mathbf k}$.  We retain the zero-th and first order contributions  in the extrinsic SOC parameter $\alpha'$. Terms of second order in $\alpha'$ are retained only insofar as they are responsible for the EY spin relaxation process.   The details of the calculation are given in the Appendix~\ref{APP_DDE}.  The final equation of motion is~\footnote{We retain the leading term in each matrix element here, as well as the current matrices below.}
\begin{equation}\label{DiffusionEquationMatrixForm}
%{\cal D}(\qv) = 
\partial_t \left(\begin{array}{c}
\Delta N_\qv\\
S^x_\qv\\
S^y_\qv\\
S^z_\qv\end{array}\right) = -
\left(\begin{array}{cccc}
Dq^{2}-i\mathbf{q}\cdot\mathbf{v} & -i\theta_{{\rm SH}}Dq_{y}q_1 & -i\theta_{{\rm SH}}Dq_{x}q_2 & -i\theta_{{\rm SH}}(\mathbf{v}\times\mathbf{q})_{z}\\
-\theta_{{\rm SH}}(v_{y}+iDq_{y})q_1 & Dq^{2}-i\mathbf{q}\cdot\mathbf{v}+\frac{1}{\tau_{sx}} &  i\kappa(\mathbf{v}\times\mathbf{q})_{z} & (i2Dq_{x}+v_{x})q_1\\
-\theta_{{\rm SH}}(v_{x}+iDq_{x})q_2 & -i\kappa(\mathbf{v}\times\mathbf{q})_{z} & Dq^{2}-i\mathbf{q}\cdot\mathbf{v}+\frac{1}{\tau_{sy}} & -(i2Dq_{y}+v_{y})q_2\\
-i\theta_{{\rm SH}}(\mathbf{v}\times\mathbf{q})_{z} & -(i2Dq_{x}+v_{x})q_1 & (i2Dq_{y}+v_{y})q_2 & Dq^{2}-i\mathbf{q}\cdot\mathbf{v}+\frac{1}{\tau_{sz}}
\end{array}\right) \left(\begin{array}{c}
\Delta N_\qv\\
S^x_\qv\\
S^y_\qv\\
S^z_\qv\end{array}\right)\,,
\end{equation}
\end{widetext}
with $q_1=2m\lambda_1$,  $q_2=2m\lambda_2$, and $k_F$ the Fermi wave vector. 
The dimensionless parameter $\kappa=\alpha' k_F^2 $ describes the efficiency of spin current swapping, which will be discussed later.  $\mathbf v=\tau e\mathbf E/m$ and $D=\tau\langle k^2/(2m^2)\rangle$ 
represent the drift velocity and the two-dimensional diffusion constant, respectively. 

 We notice that the coupling between charge and spin degrees of freedom is controlled by a cumulative spin Hall angle, $\theta_{\rm SH}=\theta_{\rm SH}^{\rm ss}+\theta_{\rm SH}^{\rm sj}+\theta_{\rm SH}^{\rm int}$, which sums up the contributions due to skew scattering, side-jump and intrinsic mechanisms:
 \ber
 \theta_{\rm SH}^{\rm ss}&=&\frac{\alpha'n_i}{2\pi} \left(\frac{mv_0}{\hbar^2}\right)^3\frac{m D}{\hbar}\,,\label{ThetaSS}\\ 
 \theta_{\rm SH}^{\rm sj}&=&-\frac{2\alpha'm}{\hbar \tau}\,, \label{ThetaSJ}\\
\theta_{\rm SH}^{\rm int}&=&\frac{2\lambda_1 \lambda_2 m \tau}{\hbar}\,,\label{ThetaInt}
\eer
%where $n_i$ is the two-dimensional density of impurities and $u$ is the strength of the short-range electron-impurity potential
 (we have reinstated $\hbar$ to highlight the dimensionless character of the spin Hall angle).
However, this charge-spin coupling is asymmetric, due to the presence of the electric field, which manifests itself in the drift velocities $v_x$ and $v_y$ in the first column of the matrix.   (To avoid misunderstanding we point out that the spin Hall angle controls only part of the total spin Hall current: the complete spin Hall current also contains a diffusion term---see next section---which is responsible for the well-known vanishing of the spin Hall conductivity in the absence of spin-orbit coupling 
from
%to 
impurities.) The spin relaxation times arise from the combination of the DP and EY mechanisms:
%The spin relaxation times arise from the combination of the D'yakonov-Perel' (DP) and the Elliot-Yafet (EY) mechanisms:
\be\label{FullSpinRelaxationTime}
 1/\tau_{si}=1/\tau_{si}^{\rm DP}+1/\tau_{si}^{\rm EY}\,,~~~~(i=x,y,z)\,,
 \ee
 where, for the special case of a (001) quantum well of a zincblende semiconductor,
\ber\label{SpinRelaxationTimes}
{1}/{\tau^{\rm DP}_{sx}}&=&Dq_1^2\,,\nn\\ 
{1}/{\tau^{\rm DP}_{sy}}&=&Dq_2^2\,, \nn\\
{1}/{\tau^{\rm DP}_{sz}}&=&D(q_1^2+q_2^2)\,.
\eer
and
\ber
{1}/{\tau^{\rm EY}_{sx}}={1}/{\tau^{\rm EY}_{sy}}=(\alpha'k_{F}^{2})^{2}/\tau \,.
\eer
The vanishing of  ${1}/{\tau^{\rm EY}_{sz}}$ is a somewhat artificial feature of our model, in which we have assumed the impurity potential to be strictly two-dimensional  and thus conserving the $z$-component of the spin.  A more realistic model, in which the impurity potential depends also on $z$, would yield finite EY relaxation time in the $z$ direction.~\cite{Averkiev08_EY,Jiang09}
%The finite EY relaxation rates along the other two directions are identical:Note that the impurity potential in the realistic quantum well may not be strictly two dimensional, where the EY spin relaxation along $z$ direction is possible.~\cite{Averkiev08_EY,Jiang09}}

When $\theta_{\rm SH}=\theta_{\rm SH}^{\rm int}$ and $\Ev=0$, the above Eqs.~(\ref{DiffusionEquationMatrixForm})  reduce to those of Ref.~\onlinecite{Raimondi06}.  
%R. Raimondi, C. Gorini, P. Schwab, M. Dzierzawa PRB {\bf 74}, 035340 (2006) [cf. Eqs. (A15) and (B1)].~\cite{Raimondi06}   
We have thus generalized those equations to take into account not only the extrinsic mechanisms of spin Hall effect, but also the effect of the electric field.   

\subsection{Equations for the currents}\label{ECurrent}
The diffusion equations derived in the previous subsection correspond to a ``reduction" of the full set of equations~(\ref{eqs1})-(\ref{eqsc}),  amounting to an elimination of the currents in favor of the densities.   To complete the formalism we must now derive the expressions for  the charge  and spin current  densities.  The ``obvious" expression  ${\tilde {\bf J}}^i=\sum_{\mathbf k}(1/2){\rm Tr}[\rho_{\bf k}\{\sigma^i,\nabla_{\bf k} H_\kv\}]$  is  incomplete, because it fails to include the anomalous velocity arising from the spin-orbit coupling with impurities.   The complete and correct expression for the matrix element of the velocity between states $\kv$ and $\kv'$ is
\ber\label{AnomalousVelocity}
\mathbf v_{\mathbf k\mathbf k'} &=&(\nabla_{\bf k} H_\kv) \delta_{\kv\kv'}- i[H_U,\mathbf r]_{\kv\kv'}\nn\\
&=&
(\nabla_{\bf k} H_\kv) \delta_{\kv\kv'}-i\alpha'\bg \sigma\times(\mathbf{k}-\mathbf{k}')v_0\,.
\eer
The current due to the last term on the right hand side can be calculated from the 
%zero-th order
%{zero-th order in spin-orbit coupling and first order in the impurity potential} approximation for the ``off-diagonal" density matrix $\rho_{\mathbf k'\mathbf k}\simeq i\pi v_0\delta(\epsilon_{\mathbf{k}}-\epsilon_{\mathbf{k}'})(\rho_{\mathbf{k}'}-\rho_{\mathbf{k}})$.  After some simplification,
``off-diagonal" density matrix $\rho_{\mathbf k'\mathbf k}\simeq i\pi v_0\delta(\epsilon_{\mathbf{k}}-\epsilon_{\mathbf{k}'})(\rho_{\mathbf{k}'}-\rho_{\mathbf{k}})$, which is zero-th order in spin-orbit coupling and first order in the impurity potential.  After some simplification,
%with 
%the relaxation time approximation, 
the corresponding contribution is cast in the 
relaxation time approximation
form
\be
{\tilde {\bf J}^{\prime i}}=\frac{\alpha'}{\tau}\sum_{\mathbf k}(1/2){\rm Tr}[\rho_{\bf k}\{\sigma^i, \kv \times \sigma\}]
\ee
%By taking into account the impurity-induced anomalous velocity $\mathbf v_{\mathbf k\mathbf k'}=-i[H_U,\mathbf r]=-i\alpha'\bg \sigma\times(\mathbf{k}-\mathbf{k}')v_0$ and zeroth order correlation between different momentum in the density matrix $\rho_{\mathbf k'\mathbf k}\simeq i\pi v_0\delta(\epsilon_{\mathbf{k}}-\epsilon_{\mathbf{k}'})(\rho_{\mathbf{k}'}-\rho_{\mathbf{k}})$, the correction in the spin current is given by ${\tilde {\bf J}^{\prime i}}=\sum_{\mathbf k \mathbf k'}(1/2){\rm Tr}[\sigma^i(\rho_{\mathbf k\mathbf k'}{\bf v}_{\mathbf k'\mathbf k}+{\bf v}_{\mathbf k\mathbf k'}{\rho}_{\mathbf k'\mathbf k})]$. After some simplification with relaxation time approximation, the total charge and spin currents can be expressed as
and the complete current is
\begin{equation}\label{J_tot}
  {{\bf J}}^i={\tilde {\bf J}^{i}}+{\tilde {\bf J}^{\prime i}}=\sum_{\mathbf k}(1/2){\rm Tr}[\rho_{\bf k}\{\sigma^i, {\tilde{\bf v}_{\bf k}}\}]
\end{equation}
with a modified velocity operator $\tilde {\bf v}_{\bf k}=\nabla_{\kv}H_\kv+(\alpha'/\tau)\mathbf k\times \bg \sigma$. We notice that this result is consistent with the calculation of the velocity from the time derivative of the ``physical" position operator, discussed, for example,  in Ref.~\onlinecite{Hankiewicz06a,Hankiewicz06b,Culcer10,Bi13}.  Finally, by substituting in Eq.~(\ref{J_tot})  the solution for the density matrix $\rho_\kv$, we obtain
the currents in terms of the densities
\begin{widetext}
\begin{equation}
\left(\begin{array}{c}
J_x^0(\mathbf q)\\
J^x_x(\mathbf q)\\
J^y_x(\mathbf q)\\
J^z_x(\mathbf q)\end{array}\right)= 
\left(\begin{array}{cccc}
-(iDq_{x}+v_{x}) & 0 & -\theta_{{\rm SH}}Dq_2 & \theta_{{\rm SH}} (iDq_{y}+v_{y})\\
0 & -(iDq_{x}+v_{x}) & i\kappa Dq_y & Dq_1\\
0 & -i\kappa Dq_y  & -(iDq_{x}+v_{x}) & 0\\
\theta_{{\rm SH}} (iDq_{y}+v_{y}) & -Dq_1 & 0 & -(iDq_{x}+v_{x})
\end{array}\right)
\left(\begin{array}{c}
\Delta N_{\mathbf q}\\
S_{\mathbf q}^x\\
S_{\mathbf q}^y\\
S_{\mathbf q}^z\end{array}\right),
\label{J_x}
\end{equation}
\begin{equation}
\left(\begin{array}{c}
J_y^0(\mathbf q)\\
J^x_y(\mathbf q)\\
J^y_y(\mathbf q)\\
J^z_y(\mathbf q)\end{array}\right)= 
\left(\begin{array}{cccc}
-(iDq_{y}+v_{y}) & -\theta_{{\rm SH}} Dq_1 & 0 & -\theta_{{\rm SH}} (iDq_{x}+v_{x})\\
0 & -(iDq_{y}+v_{y}) & -i\kappa Dq_x & 0\\
0 & i\kappa Dq_x & -(iDq_{y}+v_{y}) & -Dq_2\\
-\theta_{{\rm SH}} (iDq_{x}+v_{x}) & 0 & Dq_2 & -(iDq_{y}+v_{y})
\end{array}\right)\left(\begin{array}{c}
\Delta N_{\mathbf q}\\
S_{\mathbf q}^x\\
S_{\mathbf q}^y\\
S_{\mathbf q}^z\end{array}\right).
\label{J_y}
\end{equation}
\end{widetext}

The details of the derivation can be found in Appendix~\ref{APP_current}. Again, the coupling between charge current and spin polarization, as well as that between spin current and charge density, is found to be proportional to the total spin Hall angle $\theta_{{\rm SH}}$. 

We can now verify by direct inspection that the expressions for the charge and spin currents read from Eqs.~(\ref{J_x}) and (\ref{J_y}) agree with the phenomenological equations~(\ref{eqcc}) and (\ref{eqsc}).  The presence of the spin current swapping term in Eq.~(\ref{eqsc}) is essential to obtain this perfect agreement, as will become evident in the discussion of Section~\ref{SC_swapping}.  This gives us confidence that the spin-current swapping effect has been included properly at the phenomenological level. 
Moreover, by substituting Eqs.~(\ref{J_x}) and (\ref{J_y}) into the continuity equations, i.e., Eqs.~(\ref{eqs1}) and ~(\ref{conti1}),  we can demonstrate that the resulting drift-diffusion equations for the densities coincide with Eq.~(\ref{DiffusionEquationMatrixForm}), proving the consistency of our theory.

\section{Drift-diffusion equations at  work: Homogeneous situations}
In this section we consider a few basic applications of the formalism to homogenous situations, for which the wave vector $\qv=0$.   For definiteness we consider a (001) quantum well in a semiconductor of the zincblende structure (e.g. GaAs) with Rashba and Dresselhaus spin-orbit interactions.  The $SU(2)$ vector potentials for this system are given in Eqs. (\ref{QW001}).  The effects we study are the Edelstein effect and its inverse, the spin Hall effect and its inverse, and the spin-current swapping effect.

\subsection{Edelstein effect and its inverse}
As a first application, consider the generation of a spin polarization from an electric field  applied along the $x$ direction.  From the third of Eqs.~(\ref{DiffusionEquationMatrixForm})  we find that the time evolution of $S_y$ is determined by
\begin{equation}
 \partial_t S^y =2m \lambda_2 \theta_{\rm SH} v_x N-\frac{S^y}{\tau_{sy}}
 %\left(\frac{1}{\tau_{sy}^{\rm DP}}+\frac{1}{\tau_{sy}^{\rm EY}}\right)\,.
  \label{SXK}
\end{equation}
where the first term on the right-hand side is the spin-pumping generated by the partial conversion of the charge current into a transverse spin current, while the second term represents the spin relaxation process.  In the steady state, setting the time derivative of $S^y$ to zero, one obtains the spin density 
\begin{equation}\label{EE_py}
S^y=2\theta_{\rm SH}m \lambda_2 \tau_{sy} J_x^0\,, 
%$\frac{\tau_{sy}^{\rm DP}}{1+(\tau_{sy}^{\rm DP}/\tau_{sy}^{\rm EY})}.
\end{equation}
where we have used $v_xN= J_x^0$ to zero-th order in the SOC.
%When the spin relaxation for $S^y$ component is dominated by the DP mechanism, corresponding to $\tau_{sy}^{\rm DP}\ll \tau_{sy}^{\rm EY}$, the spin polarization reduces to ${\theta_{\rm SH}v_x}/(2mD\lambda_2 )$. In the opposite limit, the EY mechanism dominates the $S^y$ spin relaxation. The spin polarization in the steady state is expressed as $2\theta_{\rm SH}m\lambda_2 v_x \tau^{\rm EY}_{sy}$, which is proportional to the spin relaxation time due to the EY mechanism.  
As expected, the spin polarization vanishes for $\lambda_2\to0$, which corresponds to weak band SOC limit or balanced Dresselhaus and Rashba SOCs.

For the inverse process, i.e., the charge current induced by a non-equilibrium homogeneous spin-accumulation $S^y$,  the first  of Eqs.~(\ref{J_x}) gives
\begin{equation}
J_x^0=-v_x N-2\theta_{\rm SH} Dm\lambda_2  S^y\,.
\end{equation}
In the absence of an electric field ($v_x=0$) we recover the known expression for the inverse Edelstein effect:~\cite{ShenVR14}
\be\label{IEE}
J_x^0=-2\theta_{\rm SH} Dm\lambda_2 S_y \,.
\ee

We note that the role of ``driving field" in the direct Edelstein effect is played by the electric field $E_x$, while in the inverse Edelstein effect it is played by the ``spin injection field" $\dot B^y$ (see Ref.~\onlinecite{ShenVR14}).    Thus, to check  Onsager's reciprocity relations  we must compare the ratio  $S^y/E_x$ from Eq.~(\ref{EE_py}) to the  ratio  $J_x^0/\dot B^y$ from Eq.~(\ref{IEE}).  
Substituting  $J_x^0 = NE_x \tau/m$ in Eq.~(\ref{EE_py}), and $S_y = -N_0\dot B^y \tau_{sy}$~(Ref.~\onlinecite{ShenVR14}) in Eq.~(\ref{IEE}), where $N_0=N/E_F$ is the two-dimensional density of states, we can readily verify that $S^y/E_x = J_x^0/\dot B^y$, showing that Onsager's reciprocity relation is fulfilled.

\subsection{Spin Hall effect and its inverse}
Next we consider the homogeneous spin Hall effect resulting from an electric field  applied along the $x$ direction.
%It is important to point out that the spin current for spin pumping in Eq.~(\ref{SXK}) is the bare one, which is not the one measured in steady state. The latter, in our theory, can be directly calculated from the definition
According to the last of Eqs.~(\ref{J_y})  the transverse spin current is given by
\begin{equation}
J_y^z=-\theta_{\rm SH} v_xN+2m\lambda_{2}DS^{y}.
\end{equation}
We can clearly see that, as anticipated in the previous section after Eq.~(\ref{SpinRelaxationTimes}), the spin Hall angle $\theta_{\rm SH}$ controls only part of the spin current---a part that we  refer to as ``drift current".   
The remaining part is a ``diffusion current", which arises in the $SU(2)$ theory even in the absence of a spatial gradient of the spin density.  Its physical origin is in the spin precession caused by the Rashba and Dresselhaus fields.   By substituting the steady state solution of spin polarization, i.e., Eq.~(\ref{EE_py}), we arrive at the complete expression for the spin Hall spin current: 
\begin{equation}\label{SpinHallEffect}
  J_y^z=-\frac{\theta_{\rm SH}}{({\tau_{sy}^{\rm EY}}/{\tau_{sy}^{\rm DP}})+1}J_x^0,
\end{equation}
(again,  we have used $v_xN= J_x^0$ to zero-th order in the SOC) which correctly describes the crossover between the finite impurity-driven spin Hall conductivity in the limit of weak spin precession (${\tau_{sy}^{\rm DP}}\gg {\tau_{sy}^{\rm EY}}$) and the vanishing spin Hall conductivity in the strong precession  (${\tau_{sy}^{\rm DP}}\ll {\tau_{sy}^{\rm EY}}$).~\cite{Raimondi_AnnPhys12}

To exhibit the inverse spin Hall effect, we consider the first  of Eqs.~(\ref{J_y}), which yields, in the absence of an electric field  in the $y$ direction,
% the charge current along $y$ direction in response to the spin polarization, i.e.,
\be\label{JY01}
J_y^0=-2\theta_{\rm SH}Dm\lambda_1 S^x-\theta_{\rm SH}v_x S^z\,.
\ee
The first term on the right-hand side is the inverse Edelstein effect current.  The
 second term, in the leading order, can be written as $-\theta_{\rm SH}J_x^z$, where $J_x^z$ is the spin current.
We now observe, according to the second of Eqs.~(\ref{DiffusionEquationMatrixForm}), at the steady state
\be
\frac{S^x}{\tau_{sx}} +2 m\lambda_1 v_xS^z=0\,.
\ee
We can then write
\be
S^x = -2m\lambda_1 \tau_{sx} J_x^z
\ee
and, by substituting in Eq.~(\ref{JY01}), get
\ber
J_y^0&=&[D(2m^2\lambda_1)^2\tau_{sx}-1]\theta_{\rm SH}J_x^z\nn\\
&=&\left(\frac{\tau_{sx}}{\tau^{\rm DP}_x}-1\right)\theta_{\rm SH}J_x^z\nn\\
&=&-\frac{\theta_{\rm SH}}{({\tau_{sy}^{\rm EY}}/{\tau_{sy}^{\rm DP}})+1}J_x^z\,.
\eer
This equation is the mathematical formulation of the inverse spin Hall effect, which converts a spin current into a perpendicular charge current.  Comparison with Eq.~(\ref{SpinHallEffect}) shows that the Onsager reciprocity relation is satisfied. 

\subsection {Spin current swapping}\label{SC_swapping}
As a final example in the homogeneous class we discuss the spin-current swapping (SCS) effect~\cite{Lifshits09}, whereby a primary spin current,  $[J_j^i]^{(0)}$, induces a transverse spin current in which the spin direction and the direction of flow are interchanged according to the equation
 \be\label{ISCS}
[J_i^j]^{\rm SCS}=\kappa \left([J_j^i]^{(0)}- \delta_{ij}[J_l^l]^{(0)} \right)\,.
\ee

To observe experimentally the spin swapping effect, one might think to apply a uniform electric field $E_x$ to a homogeneous electron liquid, with a uniform  spin-polarization $S^x$  in the $x$ direction.  This naturally creates a primary spin current $[J_x^x]^{(0)}=-v_x S^x$, which should then induce the spin current
\begin{eqnarray}
  {[J_y^y]}^{\rm SCS}&=&-\kappa [J_x^x]^{(0)}=\kappa v_x S^x\,.\label{scsyy}
%\\
%  {[J_y^x]}^{\rm SCS}&=&\kappa [J_x^y]^{(0)}.\label{scsxx}
\end{eqnarray} 
%where $[J_j^i]^{(0)}$ and $[J_i^j]^{\rm SCS}$ represent the primary spin current and the  ``swapped" spin current, respectively.
%However, we should point out that the modality of spin current injection is critically important to the observation of spin current swapping. 
%For a homogeneously spin polarized system driven by an external electric-field in $x$ direction, the primary spin current can be expressed as $[J_x^x]^{(0)}=-v_x S^x$ and $[J_x^y]^{(0)}=-v_x S^y$, which immediately leads to
%\begin{eqnarray}
%{[J_y^y]}^{\rm SCS}&=&\kappa v_x S^x,\label{jyy1}\\
%{[J_y^x]}^{\rm SCS}&=&-\kappa v_x S^y.\label{jyx1}
%\end{eqnarray}
%One may then expect to observe the spin swapping current by detecting the spin current flowing in $y$ direction. 
Unfortunately,  our equations demonstrate  that the swapped spin current  is undetectable in  this homogeneous setup, because it is exactly cancelled by the $SU(2)$ diffusion current arising from the spin precession in the spin-orbit field generated by the electric field $E_x$.    This cancellation is already evident from the fact that the matrix elements containing $\kappa$  in Eqs.~(\ref{J_x}) and (\ref{J_y}) vanish in a homogeneous situation, because $\qv=0$.   However, our phenomenological Eq.~(\ref{eqsc}) gives more insight into the  underlying physics.   
%The point is that the impurity-induced spin swapping is not the whole story.  
%of transverse spin current $J_y^x$ and $J_y^y$: since 
The SOC effective magnetic field due to the in-plane external electric field is in the same direction ($z$ direction) as that from the impurity and both contribute to the SCS.   The SCS term on the right hand side of Eq.~(\ref{eqsc}) takes into account only the effect of the impurity, which, in this case, is given by Eq.~(\ref{scsyy}).  The additional effect of the electric field,
due to the vector potential $A^z_y$ in Eq.(\ref{QW001}),
 is taken into account  by the covariant derivative in Eq.~(\ref{eqsc}).  The two contributions cancel each other exactly for essentially the same reasons that lead to the cancellation (on the average)  of the force exerted by the electric field against the force exerted by the impurities on the electrons in a steady state situation.   

% by using the SU(2) gauge field $A_y^z=-m\alpha' eE_x$, which gives ${[J_y^a]}^{\rm E}=2D\epsilon^{azc}A_y^z S^c$. Specifically, the non-zero corrections are
%\begin{eqnarray}
%  {[J_y^y]}^{\rm E}&=&2D\epsilon^{yzx}A_y^z S^x=-\kappa v_x S^x,\\
%%  {[J_y^y]}^{\rm E}&=&2D\epsilon^{yzx}A_y^z S^x=-2D\epsilon^{yzx}m\alpha'eE_x S^x,\\
%  {[J_y^x]}^{\rm E}&=&2D\epsilon^{xzy}A_y^z S^y=\kappa v_x S^y.
%\end{eqnarray} 
%which completely cancel the spin currents due to spin swapping term in Eqs.~(\ref{jyy1}) and (\ref{jyx1}). 
In order to observe the spin swapping effect in an experiment, one should avoid the influence from electric-field-induced SOC.  One way to achieve this  is to inject the spin current by optical means.  Another possibility is to inject a  pure spin current via the spin Seebeck effect~\cite{SSeebeck} or  via multi-terminal electrical spin injection techniques.~\cite{Lou07}
We will return to this point in Section~\ref{SCSgradient}.

\section{{Drift-diffusion equations at  work: inhomogeneous situations}}
Let us now consider some applications of our theory to inhomogeneous situations.  We have in mind, specifically, the electron-hole density waves and spin density waves which can be optically induced on the surface of a semiconductor quantum well through the interference of laser beams coming from different directions with different polarizations.   These structures are also referred to as ``gratings", because the non-uniformity of the densities causes a modulation in the refractive index of the electron gas.  The spin density generated in this manner is typically associated with the electrons only (the holes losing their polarization during a very short relaxation time) and is perpendicular to the plane of the quantum well.  Recently developed pump-probe techniques have allowed detailed studies of the spontaneous dynamical evolution of these systems on a picosecond time scale:  by this we mean that it is possible to record the density, the spin density and the overall velocity of propagation of the grating on a picosecond scale.  Through such experiments it has been possible, for example, to establish the presence of a long-lived ``persistent spin helix" in nearly balanced  (001) quantum wells  (i.e., quantum wells with $\alpha\simeq \pm \beta$), and to demonstrate interesting effects related to Coulomb drag in spin diffusion and electron-hole diffusion.  
%\begin{figure}
%\includegraphics[width=6cm]{balancedSOC.pdf}
%\caption{(Color online) Effective magnetic fields at Fermi surfaces in two balanced SOC configurations with grating wave vectors along $x$ (left) and $y$ (right) axises, respectively.}
%\label{figbSOC}
%\end{figure}
In this section we focus on the partial conversion of an electron-hole density grating into an electronic spin density wave under the action of an electric field which we choose, for definiteness, to be  parallel to the $x$ axis: $\Ev=E{\bf \hat x}$.  The numerical calculations  are carried out for a 10~nm GaAs QW grown in (001) direction, in which the Dresselhaus coefficient $\beta=\unit{10}{meV\cdot\AA}$, unless otherwise specified. 
The Rashba coefficient, $\alpha$,  is assumed to be tunable via gate voltage. The Elliott-Yafet spin-relaxation process is neglected.
\begin{figure}
\includegraphics[width=6cm]{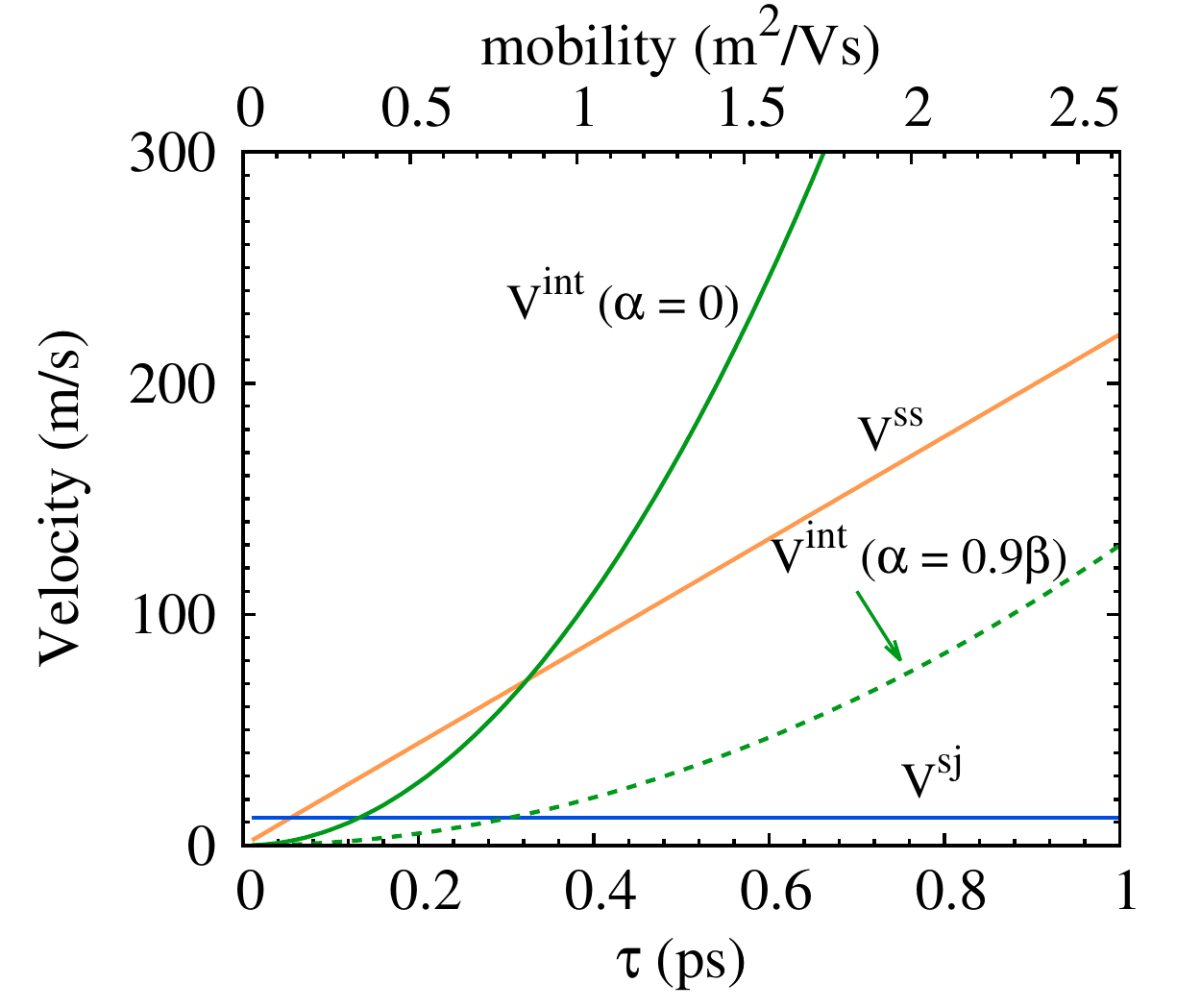}
\caption{(Color online) Magnitudes of spin Hall velocities due to skew scattering ($v^{\rm ss}$), side-jump ($v^{\rm sj}$) and intrinsic mechanism ($v^{\rm int}$) as functions of momentum relaxation time (lower axis) and mobility (upper axis) in 10~nm GaAs QW grown in (001) direction with electron density $n_e=\unit{10^{12}}{cm^{-2}}$. The electric field is taken to be $\unit{1}{kV/cm}$. The results for the intrinsic mechanism are computed with two different values of Rashba coefficient.}
\label{fig2}
\end{figure}

In Fig.~\ref{fig2}, we plot the magnitudes of the three components of the homogeneous spin Hall drift velocity $v^{\rm ss}=\theta_{\rm SH}^{\rm ss}v$, $v^{\rm sj}=\theta_{\rm SH}^{\rm sj}v$, and $v^{\rm int}=\theta_{\rm SH}^{\rm int}v$ as functions of momentum relaxation time.  The latter is varied by changing the concentration of impurities.  % {\bf (Note that the sign of the spin Hall velocity due to side-jump mechanism is opposite to those due to skew scattering and intrinsic mechanisms, according to the definitions of spin Hall angles Eqs.~(\ref{ThetaSS})-(\ref{ThetaInt}))} 
 Here $v$ is the standard drift velocity of the electrons given by $v=\mu E$, where $\mu\simeq e\tau/m$ is the mobility, whose value is shown on the upper axis of Fig.~\ref{fig2}.  Qualitatively, the side jump drift velocity is independent of momentum relaxation time, whereas the intrinsic spin Hall drift velocity and the skew scattering spin Hall drift velocity  are proportional to $\tau^2$ (provided $\tau \ll \tau_{DP}$) and $\tau$ respectively.  It is seen that the spin Hall drift velocities due to the side jump effect and the skew scattering are  comparable when the mobility is below $\unit{0.1}{m^2V^{-1}s^{-1}}$, while the skew scattering is dominant in an intermediate regime. In the high mobility region, the relative contribution of skew scattering and intrinsic mechanism can be effectively controlled by changing the Rashba coefficient. This can be easily understood from $v^{\rm int}=2\tau m (\beta^2-\alpha^2)v$, which demonstrates the vanishing of the intrinsic mechanism for $\alpha=\pm\beta$.  
In the following we consider a few situations of experimental interest, in which an electric field is applied to a density or a spin density grating, parallel or perpendicular to the direction of the wave vector.

\subsection{Density grating with $\mathbf q \| \mathbf E$: Periodic Edelstein effect.}
Let us begin with the case in which the system is initially prepared in a density wave state (no spin density), with the wave vector of the electron-hole density wave parallel to the direction $x$ of the external electric field.    
%the external electric field, an initial electron-hole grating can creates spin grating with only $S^y$ spin component, because the other two spin components are completely decoupled from density grating. 
From the drift-diffusion equations~(\ref{DiffusionEquationMatrixForm}) we  see that only the $y$-component of the spin density is coupled to the particle density: 
\begin{eqnarray}
\partial_t \Delta N_\qv&=&(i{q}{v}-D_aq^{2})\Delta N_\qv+i\theta_{SH} D_sqq_2S^y_\qv,\label{nsy}\\
\partial_t S^y_\qv&=&[i{q}{v}-D_s(q^{2}+q_2^2)]S^y_\qv+ \theta_{{\rm SH}}(iD_sq+v)q_2\Delta N_\qv\,.\nonumber\\
\end{eqnarray}
  The spin-density coupling strength is proportional to $\theta_{\rm SH}q_2$,  where $q_2\equiv2 m\lambda_2=2m(\beta-\alpha)$, and therefore it vanishes in the balanced case $\alpha=\beta$.   
  %This is the collective (grating) analogue of the Edelstein effect.   
%We should point out that the couplings here, namely between density and in-plane spin polarization, actually correspond to the EE and IEE in inhomogeneous case. 
Notice that we have replaced the plain diffusion constant $D$ by the ambipolar diffusion constant $D_a$ for the electron-hole density grating and by the spin diffusion $D_s$ for the spin grating.~\cite{ShenV13b}  The first replacement takes into account the almost perfect screening of the space charge that occurs when electrons and holes diffuse together in an electron-hole density wave.~\cite{ShenV13a,ShenV13b,Yang2011a}  The second takes the effect of spin Coulomb drag, which reduces the spin diffusion constant relative to standard $D$.~\cite{Weber05,Kikkawa99,Flatte00,DAmico00,Flensberg01}  

Neglecting the feedback from spin grating to density grating, i.e., the second term on the right-hand side in Eq.~(\ref{nsy}), which is of second order in spin-Hall angle, we find that the density evolves according to the standard analytic formula 
\be
\Delta N_\qv=A_0 e^{(iqv-D_aq^2)t},\label{DeltaN}
\ee
where  $A_0$ is the amplitude of the initial electron-hole density grating.
 Substituting this in the equation for the spin density we obtain an analytically solvable equation whose solution is
\begin{eqnarray}\label{Smp2}
  S^{y}_\qv&=&\frac{A_0\theta_{\rm SH}(iD_s q+ v)q_2
    \exp[{iq(x+vt)}]}{ D_{s}(q^{2}+q_{2}^{2})-D_aq^2}\nonumber\\
  &&\hspace{-0.5cm}\mbox{}\times \{\exp(-D_aq^2 t)-\exp[{-D_{s}(q^{2}+q_{2}^{2})t}]\}\,.
\end{eqnarray}
In the absence of the electric field ($v=0$) the density grating simply decays at a rate $D_aq^2$ determined by the ambipolar diffusion constant.   The spatially periodic diffusion current generates a spin polarization in the $y$-direction---an effect that can be viewed as the analogue of the uniform Edelstein effect except that: (i) it is spatially periodic and (ii) it is driven by a diffusion current. This polarization, starting from zero at the initial time, reaches a maximum  at time $t= \ln (D_a/D_s)/[D_aq^2-D_s(q^2+q_2^2)]$  before eventually tending to zero at long times, when the density grating disappears.
  
 When the electric field is applied, the  phase of the grating acquires a linear variation in time, corresponding to a drift with velocity $v$ in the direction of the electric field.  The $y$-component of the spin polarization also drifts with the same velocity  $v$.  Thus, the periodic Edelstein effect offers a way to generate a drifting in plane-polarized spin grating:  this would be  difficult, if not impossible to produce by optical means.  

%\begin{eqnarray}
%\partial_t \Delta N&=&(i{q}{v}-Dq^{2})\Delta N+i\theta_{SH} Dqq_2S^y\\
%\partial_t S^x&=&[i{q}{v}-D(q^{2}+q_1^2)]S^x- (i2Dq+v)q_1 S^z\\
%\partial_t S^y&=&[i{q}{v}-D(q^{2}+q_2^2)]S^y+ \theta_{{\rm SH}}(iDq+v)q_2\Delta N\\
%\partial_t S^z &=&[i{q}{v}-D(q^{2}+q_1^2+q_2^2)]S^z+(i2Dq+v)q_1S^x.\nonumber\\
%\end{eqnarray}

\subsection{Spin grating with $\mathbf q \| \mathbf E$: Helical Doppler effect}
To exhibit the dynamics of an optically created spin grating polarized along $z$ direction, we write the drift-diffusion equations in terms of the two helical modes $S^\pm_\qv=(S^x_\qv\pm iS^z_\qv)/\sqrt{2}$, 
\begin{equation}
  \partial_t S^\pm_\qv =\left[i(q\pm q_1){v}-D_s(q\pm q_1)^{2}+D_sq_1^2-\frac{1}{\tau^+_{s}}\right] S^\pm_\qv-\frac{S^\mp_\qv}{\tau_{s}^-}
\end{equation}
where ${1}/{\tau^\pm_{s}}=({1}/{\tau_{sx}}\pm{1}/{\tau_{sz}})/2$ and $q_1=2m \lambda_1$. For $|\lambda_1|\gg |\lambda_2|$, we can neglect the coupling between the two modes ($\propto 1/\tau_s^-$), which leads us to the analytic solution 
\be
S^\pm_\qv=(\pm iA_z/\sqrt 2)\exp\{iqx+i(q\pm q_1){v}t-D_s(q\pm q_1)^{2}t\}\,,
\ee
 where the two spin helical modes show  different phase evolutions, with phases $\phi_\pm=(q\pm q_1){v}t$.  Here, $A_z$ is the amplitude of the initial spin grating. We see that the ``Doppler shifts" are different for the two helical components. Recall that the helical mode $S_-$ describes the ``persistent spin helix" in the balanced case,~\cite{Bernevig06,Weber07,Weng08,Koralek09,Slipko11,Walser12,Tokatly13} when the wave vector $q$ of the grating matches the SOC wave vector $q_1$. In recent experiments by Yang {\it et al.}~\cite{Yang2011a,Yang2012} the time evolution of the spatial phase has been measured by Doppler velocimetry.  It is easy to see that, on a short time scale, when the exponential in the above equation can be linearized, the superposition of the two helical modes with comparable amplitudes leads to a global phase velocity $\dot\phi \simeq (\dot\phi_++\dot \phi_-)/2 \simeq qv$, which is the same as in Eq.~(\ref{Smp2}). However, on a long-time scale, only the long-lived mode $S_-$, which corresponds to the persistent spin helix, is relevant (assuming, of course, that $q$ is close to $q_1$)  leading to $\dot\phi \simeq \dot \phi_-\simeq (q-q_1)v$, which can have positive or  negative sign depending on the whether  $q>q_1$ or $q<q_1$ (Ref.~\onlinecite{Kleinert2007,Yang2010,WengPRB12_Doppler}).  The switching sign of the phase velocity has been experimentally observed and provides strong evidence for the existence of a long-lived spin helix in this system.~\cite{Yang2012}
%\begin{widetext}
%\begin{equation}
%{\cal D} = 
%Dq^{2}-i{q}{v}+\left(\begin{array}{cccc}
%0 & 0 & -2i\theta_{{\rm SH}}\tilde{\lambda}_{2}Dq & 0\\
%0 & \frac{1}{\tau_{sx}} & 0 & \tilde{\lambda}_{1}(i4Dq+2v)\\
%-2\theta_{{\rm SH}}\tilde{\lambda}_{2}(v+iDq) & 0 & \frac{1}{\tau_{sy}} & 0\\
%0 & -\tilde{\lambda}_{1}(i4Dq+2v) & 0 & \frac{1}{\tau_{sz}}
%\end{array}\right).
%\label{DiffusionEquationMatrixForm}
%\end{equation}
%\end{widetext}
\begin{figure}
  \includegraphics[width=5.5cm]{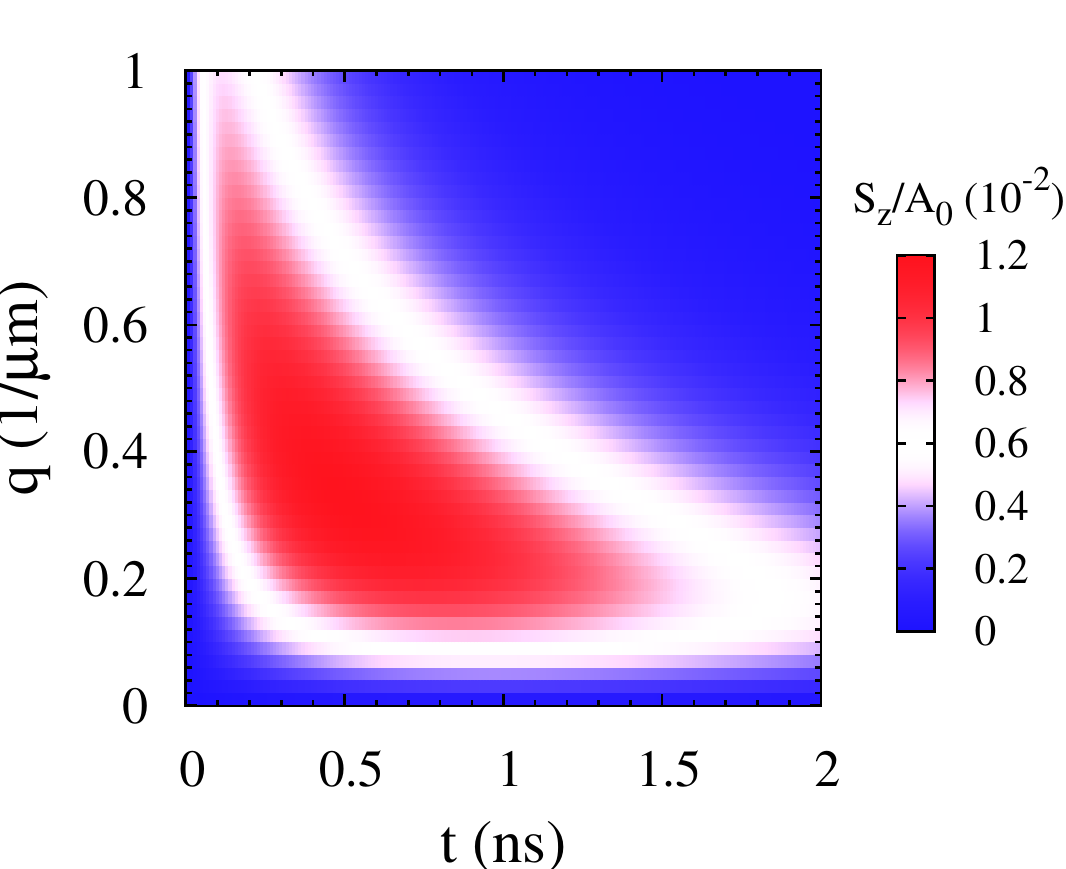}
  \caption{(Color online) Time evolution of the amplitude of the spin grating $S_z$ arising from an electron-hole grating of wave vector $q$ in the presence of an electric field perpendicular to the wave vector.   Here only the extrinsic spin Hall effect (skew scattering and side-jump effect) is considered.  The largest amplitude of the spin grating occurs in the central region of the plot (in red), and is maximized about $q$=0.2 $\mu$m$^{-1}$. In this section, we take $\tau=\unit{1}{ps}$, $D_a=\unit{20}{cm^2/s}$ and $D_s=\unit{200}{cm^2/s}$ unless otherwise specified. The electron-hole recombination rate $\Gamma=\unit{1}{ns^{-1}}$ is used.}
\label{fig3}
\end{figure}
\subsection{$\mathbf q \perp \mathbf E$: Collective spin Hall effect, extrinsic}
When $\qv$ is perpendicular to the external electric field, an electron-hole density grating becomes coupled, via the spin Hall effect, to the $z$-component of the spin density.  This coupling generates a spin-density grating polarized in the $z$-direction, which can be observed  in a  Kerr/Faraday rotation measurement. The diffusion matrix $\cal D$ in this configuration turns to be
\begin{equation}
%{\cal D} = 
\left(\begin{array}{cccc}
Dq^{2} & -i\theta_{{\rm SH}}Dqq_1 &0 & -i\theta_{{\rm SH}}{v}{q}\\
-i\theta_{{\rm SH}}Dqq_1 & D(q^{2}+q_1^2) & 0 & vq_1\\
-\theta_{{\rm SH}}vq_2 & 0 & D(q^{2}+q_2^2) & -iDqq_2\\
-i\theta_{{\rm SH}}{v}{q} & -vq_1 & 2iDqq_2 & D(q^{2}+q_1^2+q_2^2)
\end{array}\right).
\label{D_k_perp}
\end{equation}
Let us first consider  the case in which the band SOC is  zero (this is the case for a GaAs (110) quantum well, see Ref.~\onlinecite{ShenV13b}).  Then both $S^x$ and $S^y$ are decoupled from the density grating, and the solution for the $S^z$ component is given by
\begin{equation}\label{S_z}
  S^z=\frac{A_0\theta_{\rm SH}qv\sin(qx)}{D_{s}q^{2}-D_aq^2}(e^{-D_aq^2 t}-e^{-D_{s}q^{2}t})\,,
\end{equation}
with  the  spin Hall angle $\theta_{SH}$ being entirely due to electron-impurity scattering, i.e., of entirely extrinsic origin.  Here we are neglecting, for simplicity, the small electron-hole recombination rate $\Gamma$, which would  modify the decay rate of density grating from $D_a q^2$ to $D_a q^2+\Gamma$ (Ref.~\onlinecite{ShenV13b}).  A non-zero value of $\Gamma \simeq$ 1 ns$^{-1}$ is, however, included in the calculations plotted in Fig.~\ref{fig3}.   
The spin density from Eq.~(\ref{S_z}) vanishes both at short times  ($t \to 0$), when the electric field has not had sufficient time to produce its effect, and for long times ($t \to \infty$) when the original density grating has diffused away. 
Its maximum amplitude occurs at $t={\ln (D_a/D_s)}/[{(D_a-D_s)q^2}]$ and is given by
%
%By introducing the electron-hole recombination rate in Eq.~(\ref{S_z}), one can carry out the maximal amplitude of $S_z$ grating from $\partial_t S_z=0$, given by
 \begin{equation}\label{szmax}
\frac{A_{S_{z}}^{\rm max}(q)}{A_{0}}=\frac{\theta_{\rm SH}v}{D_{a}q}\left(\frac{D_{s}}{D_{a}}\right)^{{D_{s}}/({D_{a}-D_{s}})}\,,
%\frac{A_{S_{z}}^{\rm
%  max}(q)}{A_{0}}=\frac{v_{\rm ext}q}{D_{a}q^{2}+\Gamma}\left(\frac{D_{s}q^{2}}{D_{a}q^{2}+\Gamma}\right)^{{D_{s}q^{2}}/({D_{a}q^{2}+\Gamma-D_{s}q^{2}})}.
\end{equation}
which is approximately the fraction of the grating wavelength $\sim 1/q$ through which the electrons move, with drift velocity $\theta_{\rm SH}v$,  during the grating lifetime $\sim 1/D_aq^2$.   
In Fig.~\ref{fig3}, we plot the time evolution of the amplitude of the induced spin grating as a function of wave vector.   
%In these calculations, we have introduced a parameter $\Gamma$ (specified in the figure caption) to describe the electron-hole recombination, which modifies the decay rate of density grating from $D_a q^2$ to $D_a q^2+\Gamma$ (Ref.~\onlinecite{ShenV13b}). 
In this figure, the optimal value of $q$, leading to the largest amplitude of the induced spin grating, is about $q^{\rm opt}\sim \unit{0.2}{\mu m^{-1}}$, which is not too far from experimentally realized values.~\cite{Weber05,Yang2011a}

\subsection{$\mathbf q \perp \mathbf E$: Collective spin Hall effect, intrinsic}
Let us now consider the interesting case in which the collective spin Hall effect occurs in the presence of band SOC and Rashba coupling.  We will consider three cases: (i)  the SOC balanced case with $\alpha=-\beta$ ($q_1=0$ and $q_2=q_0$ with $q_0=4m\beta^2/\hbar^2$), (ii)  the SOC balanced case with $\alpha=\beta$ ($q_1=q_0$ and $q_2=0$),  and (iii) the generic $\alpha\ne \pm \beta$ case.  \\

\begin{figure}
\includegraphics[width=5.5cm]{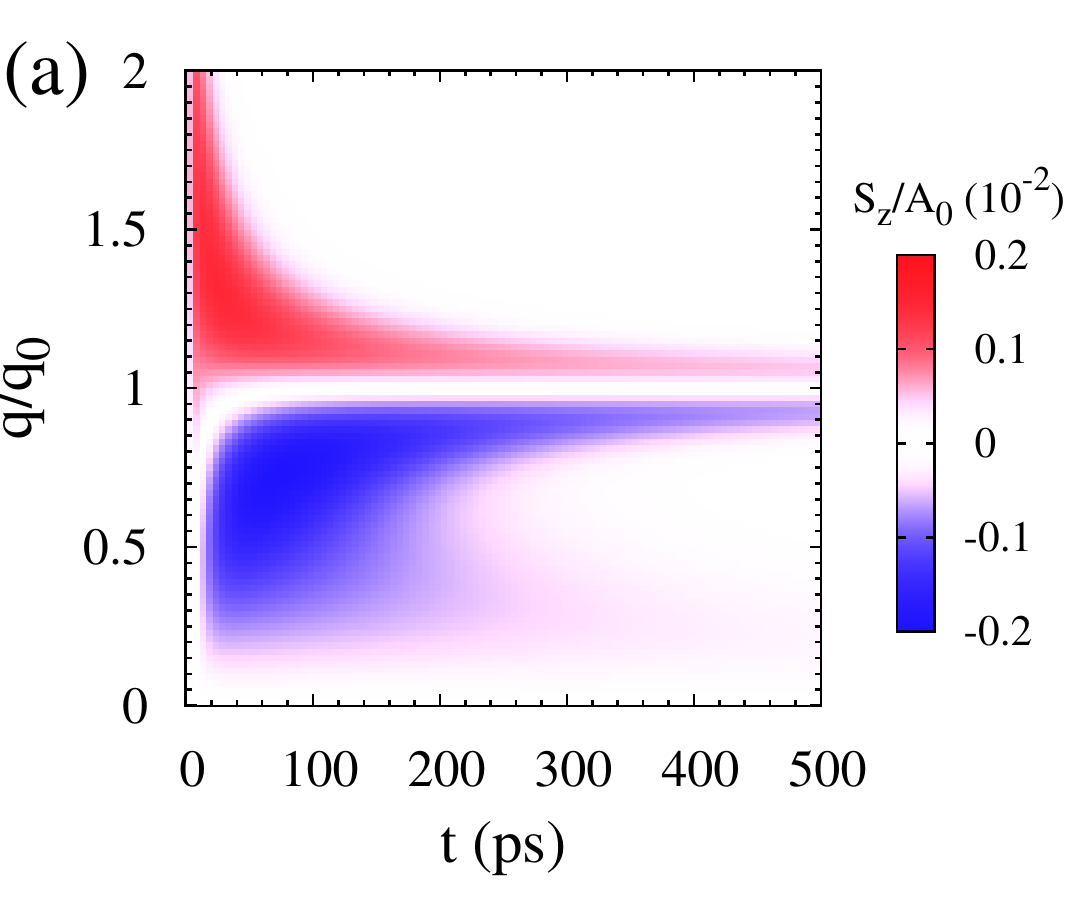}
\includegraphics[width=5.5cm]{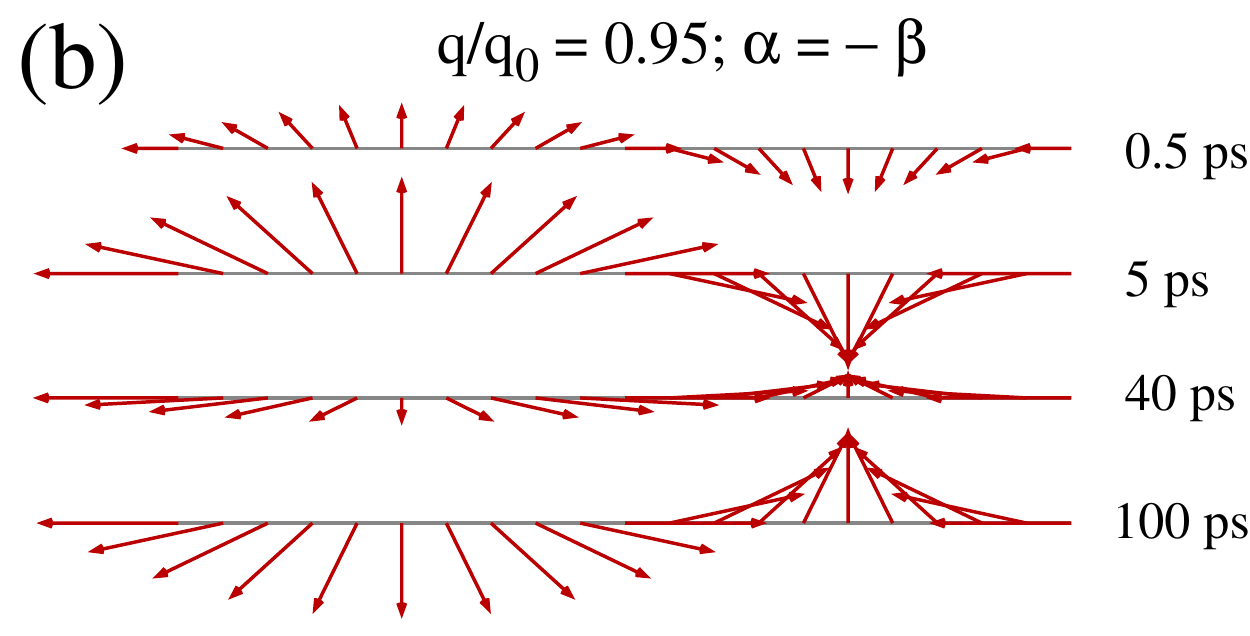}
  \includegraphics[width=5.5cm]{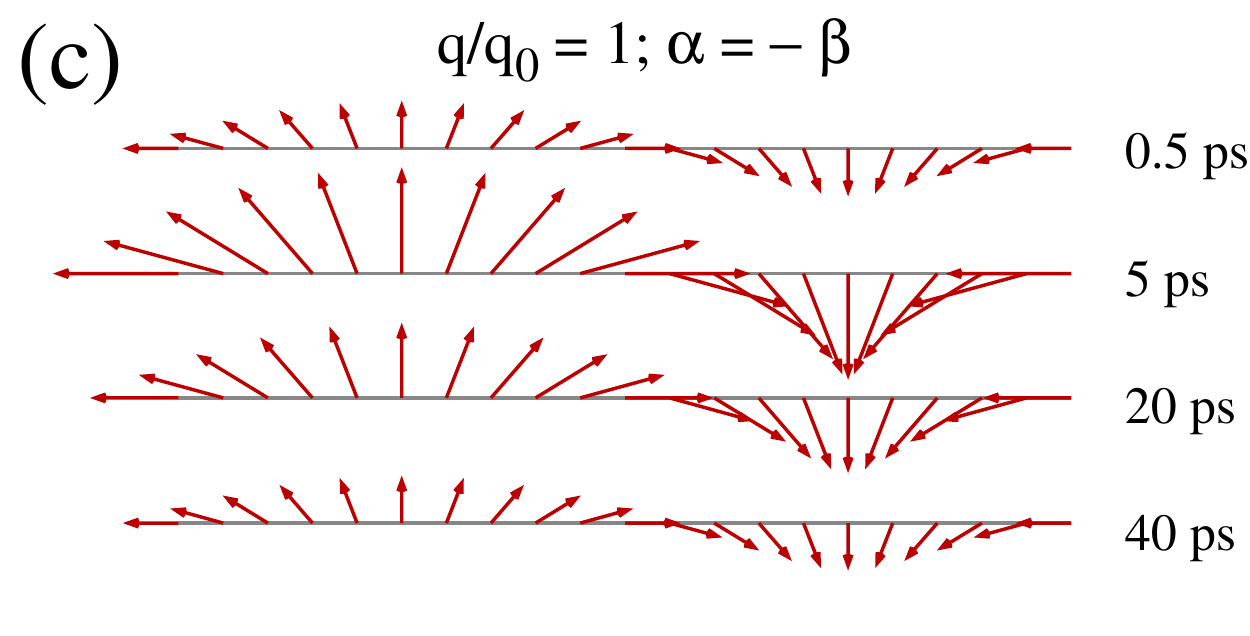}
  \includegraphics[width=5.5cm]{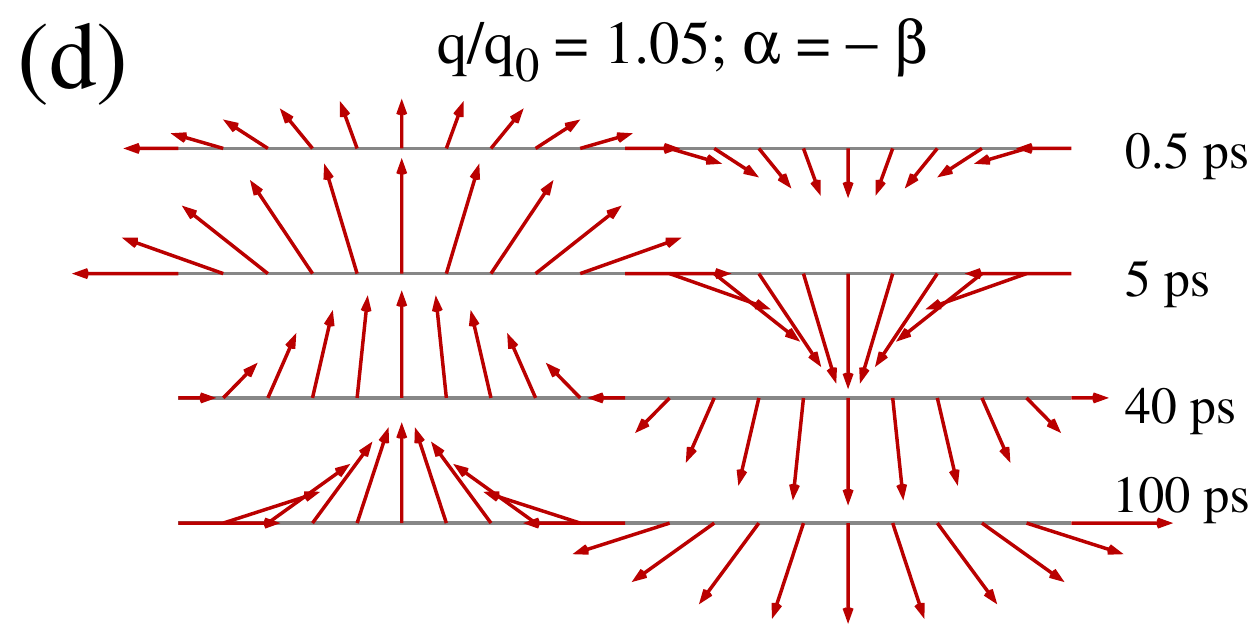}
  \caption{(Color online) (a) Time evolution of the amplitude of the spin grating $S_z$ arising from an electron-hole grating of wave vector $q$ (normalized by $q_0\simeq 3.5~\mu$m$^{-1}$) in the presence of an electric field perpendicular to the wave vector with $\alpha=-\beta$.  
 (b)-(d) The corresponding spin profiles $S^y$-$S^z$ induced by collective spin Hall effect with different grating wave vectors around $q_0$. Four typical times are chosen for each case. For $q\ne q_0$, the profile at short time is a superposition of $S^+$ and $S^-$ with weights $q-q_0$ (large) and $q+q_0$ (small). At long time, only the $S^+$ mode ($\propto q-q_0$) survives. (c) For $q=q_0$, the long lived mode disappears and the long-time spin profile is qualitatively similar to the short-time state.}
\label{fig4}
\end{figure}

 (i) In the SOC balanced case with $\alpha=-\beta$, the $x$ component of the spin  decouples from the rest, while the $S^y$ and $S^z$ components remain coupled to the density and to each other.  Transforming to the helical basis, $S^\pm=(S^y\pm iS^z)/\sqrt 2$, we obtain
\begin{eqnarray}
  \partial_t \Delta N_\qv&=&-D_aq^2(\Delta N_\qv)+\theta_{\rm SH}vq\frac{(S^+_\qv-S^-_\qv)}{\sqrt 2},
  \label{eq24}\\
  \partial_t S^+_\qv&=&(\theta_{\rm SH}/\sqrt 2)q_{-}v(\Delta N_\qv)-D_sq_{-}^2S^+_\qv,\\
  \partial_t S^-_\qv&=&(\theta_{\rm SH}/\sqrt 2)q_{+}v(\Delta N_\qv)-D_sq_{+}^2S^-_\qv,
%  \partial_t S^y&=&\theta_{\rm SH}vq_2N-D_s(q^2+q_2^2)^2S^y+D_siqq_2S^z,\\
%  \partial_t S^z&=&i\theta_{\rm SH}vqN-2iDqq_2S^y-D_s(q^2+q_2^2)^2S^z,\label{HE}
\end{eqnarray}
where $q_{\pm}\equiv q_0\pm q$.  From the last two equations, we see that the electric  field ``pumps" the spin helical modes $S^+$   and $S^-$ at a  rate proportional to $q_{-}$ and  $q_{+}$ respectively.  At the same time, the diffusion process causes these modes to decay at rates  $D_sq_{-}^2$ and  $D_sq_{+}^2$, respectively.  As before, we discard the small feedback of the spin on the evolution of the electron-hole density. Then taking the density from Eq.~(\ref{DeltaN}), but without the drift term (because $\qv \perp \Ev$), we easily obtain an analytic solution for the helical modes:
\begin{equation}
  S^{\pm}_\qv=\frac{A_0\theta_{\rm SH}vq_{\mp}
    e^{iqx}}{\sqrt 2 (D_{s}q_{\mp}^2-D_aq^2)}(e^{-D_aq^2t}-e^{-D_{s}q_{\mp}^2t}).\label{Smp}
\end{equation}
Interestingly, at the special wave vectors $q=\pm q_0$, for which a persistent spin helix is expected to appear in the $S^{+}$ channel (if $q=q_0$), or in the $S^-$ channel (if $q=-q_0$)  the present result shows that only the short-lived spin mode is generated, i.e. only $S^-$ if  $q=q_0$,  or only $S^+$ if  $q=-q_0$.  The reason for this somewhat counterintuitive behavior is that pumping the persistent helical mode is equivalent, modulo an $SU(2)$ rotation, to pumping a uniform spin polarization in the $z$-direction of the rotated frame.  But the $SU(2)$ rotation in question eliminates the band SOC, leaving only an extrinsic SOC which cannot change the spin polarization in the $z$ direction and therefore cannot ``pump" the long-lived mode.\cite{ShenV13b}    

The time evolution of the $S_z$ grating amplitude is plotted in Fig.~\ref{fig4}(a) for different magnitudes of the wave vector.  Observe the change in the sign of the amplitude of the spin grating around $q=q_0$ in the long time regime. This is because, when the grating wave vector exactly equals  $q_0$, only the short-lived chiral mode, of which the amplitude is negligibly small after 100~ps, can be pumped by spin Hall effect [also see Fig.\ref{fig4}(c), where we show the $S_z$ spin profile at different times]. For a wave vector slightly above or below $q_0$, the long-lived mode is also pumped, and inevitably becomes dominant after a few tens of picoseconds.  According to Eq.~(\ref{Smp}) the amplitude of the pumped long-lived-mode changes sign from $q>q_0$ to $q<q_0$, and vanishes at $q=q_0$.
In Figs.~\ref{fig4}(b) and ~\ref{fig4}(d) we plot the spin configurations for two typical wave vectors, $q=1.05q_0$ and $q=0.95q_0$.  At short times, the short-lived component of the response dominates, leading to similar behaviors in the two cases.   At long times, only the long-lived mode components survive, leading to responses of opposite sign. \\

\begin{figure}
\includegraphics[width=5.5cm]{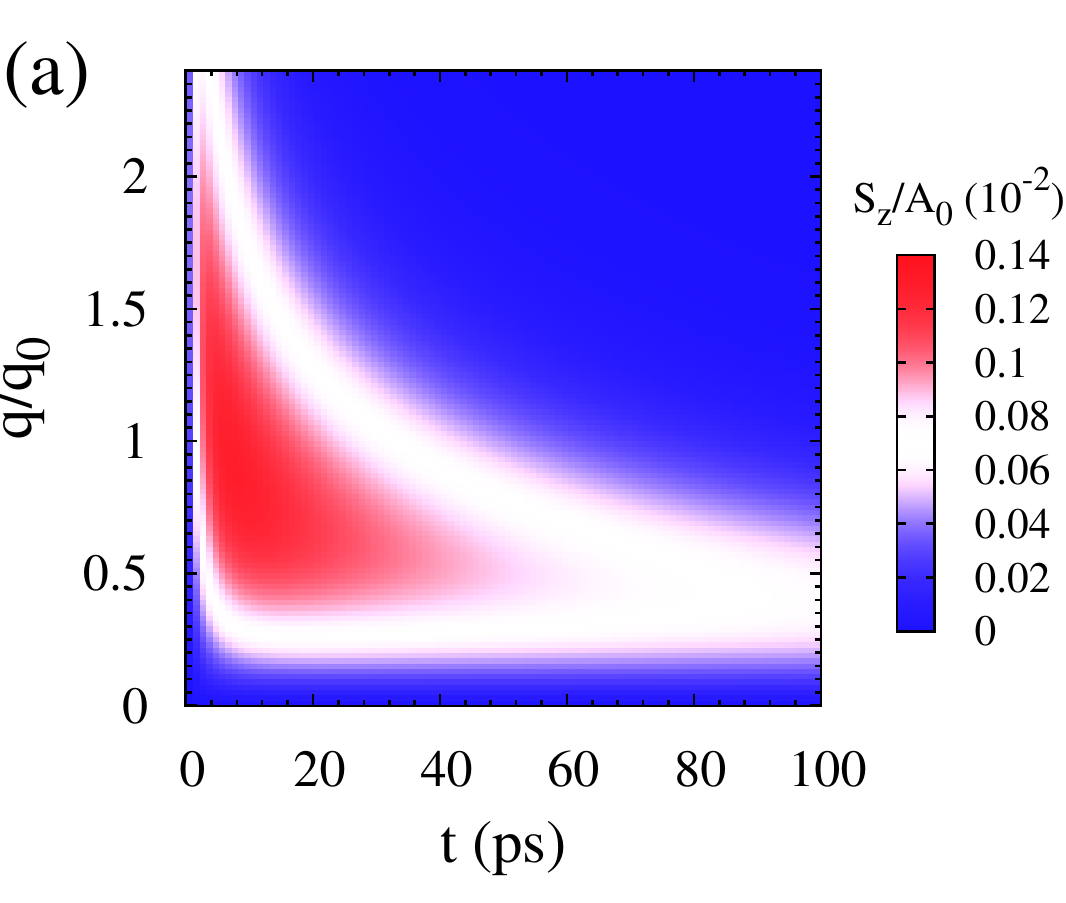}
\includegraphics[width=5.5cm]{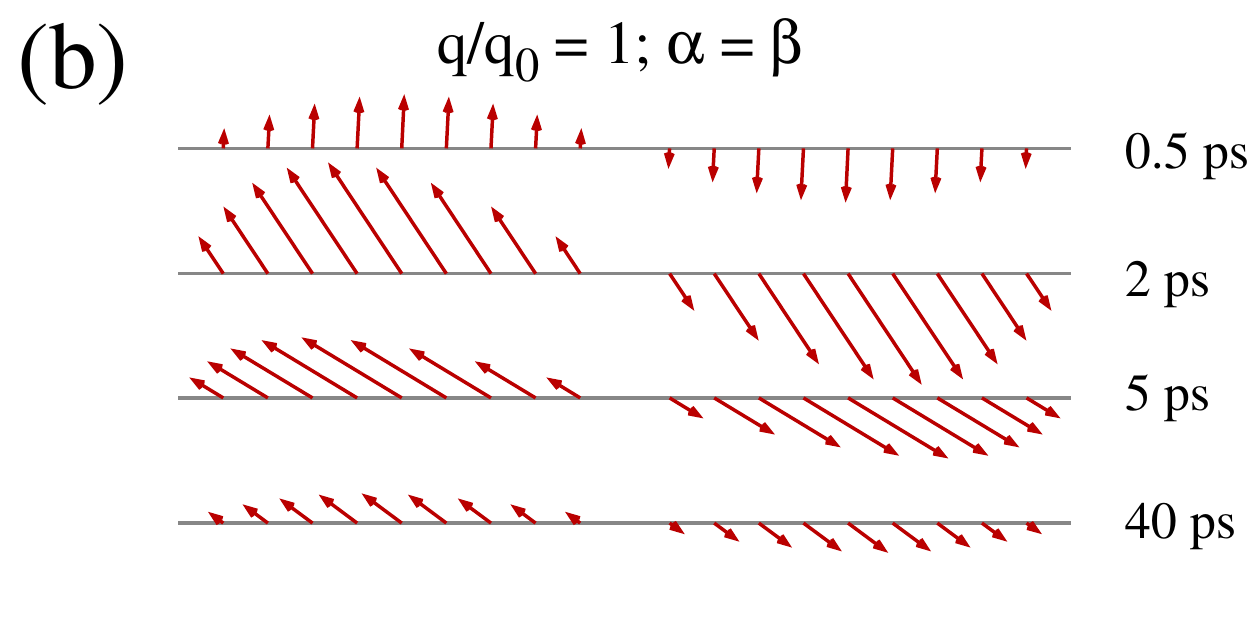}
  \caption{(Color online) (a) Time evolution of the amplitude of the spin grating $S_z$ arising from an electron-hole grating of wave vector $q$ in the presence of an electric field perpendicular to the wave vector with $\alpha=\beta$. (b) The corresponding spin profiles $S^x$-$S^z$ induced by collective spin Hall effect at $q=q_0$. We observe that the helical modes $S^+$ and $S^-$ have the same lifetime and amplitude.}
\label{fig5}
\end{figure}
(ii) In the SOC balanced case with $\alpha=\beta$, it is the coupling between $S^y$ and the other densities that vanishes. Then, the equation of motions for helical modes $S^\pm_\qv=(S^x_\qv\pm iS^z_\qv)/\sqrt 2$ takes the form
%\begin{equation}
%{\cal D} = 
%\left(\begin{array}{cccc}
%Dq^{2} & -i\theta_{{\rm SH}}Dqq_1 &0 & -i\theta_{{\rm SH}}{v}{q}\\
%-i\theta_{{\rm SH}}Dqq_1 & D(q^{2}+q_1^2) & 0 & vq_1\\
%0 & 0 & D(q^{2}) & 0\\
%-i\theta_{{\rm SH}}{v}{q} & -vq_1 &0 & D(q^{2}+q_1^2)
%\end{array}\right).
%\label{D_k_perp2}
%\end{equation}
\begin{equation}
\partial_t S^\pm_\qv=\frac{\theta_{{\rm SH}}}{\sqrt 2}(iD_sq_0\mp v)q(\Delta N_\qv) - [D_s(q^{2}+q_0^2)\mp ivq_0]S^\pm_\qv.
\end{equation}
%\begin{eqnarray}
%\partial_t S^+&=&\frac{\theta_{{\rm SH}}}{\sqrt 2}(iD_sq_1-v)q(\Delta N) - [D_s(q^{2}+q_1^2)-ivq_1]S^+,  \nonumber \\
%\partial_t S^-&=&\frac{\theta_{{\rm SH}}}{\sqrt 2}(iD_sq_1+v)q(\Delta N) - [D_s(q^{2}+q_1^2)+ivq_1]S^-. \nonumber
%\end{eqnarray}
Note that the pumping and decay rates are  the same for  the amplitudes of the two modes,  determined by $\theta_{{\rm SH}}(\Delta N)\sqrt{v^2+(D_sq_0)^2}$ and $D_s(q^2+q_0^2)$ respectively, whereas their phases are different. The time evolution of the amplitude of  the $S_z$ component is shown in Fig.~\ref{fig5}(a).
% However, the phases of the two modes are distinct, including the one produced by the pumping process and its evolution {\bf (???)}. As shown in Fig.~\ref{fig2}(d), the signal becomes very weak after first 20 ps, due to the absence of the long-live mode {\bf (???)}. 
The spin profiles with $q=q_0$ at different times are shown in Fig.~\ref{fig5}(b). \\

\begin{figure}
\includegraphics[width=5.5cm]{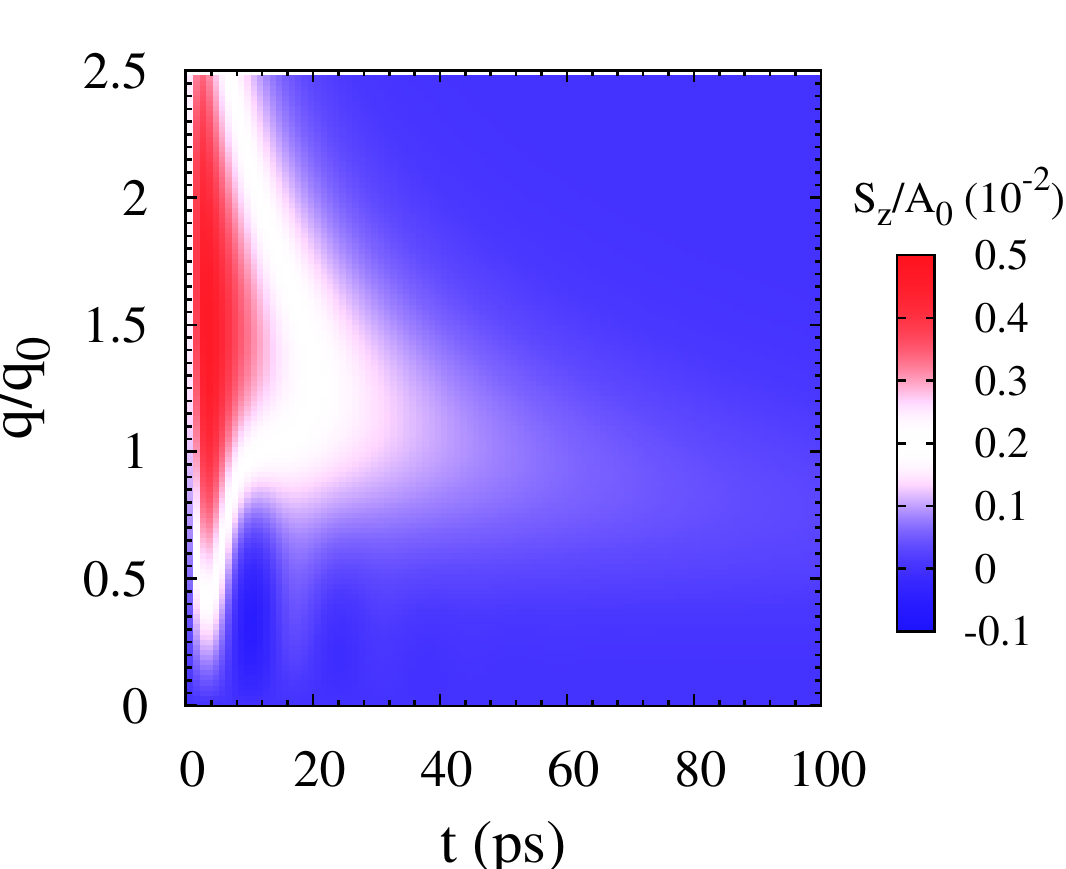}
  \caption{(Color online) Time evolution of the amplitude of the spin grating $S_z$ arising from an electron-hole grating of wave vector $q$ in the presence of an electric field perpendicular to the wave vector without Rashba SOC.}
\label{fig6}
\end{figure}

(iii) For $\alpha\ne \pm\beta$, all the spin components are coupled together, which makes the analysis much more complicated. Since the intrinsic mechanism is switched on, the collective spin Hall effect can produce large-amplitude spin gratings in a high mobility sample, see Fig.~\ref{fig6}, where the Rashba coupling is excluded. Again,  spin relaxation restricts the opportunity for observation  to a relatively short time window immediately following the initial creation of the electron-hole density grating.\\

%The time evolution of the two mode can be approximately written  as
%\begin{eqnarray}
%  S^{\pm}&=&\frac{A_0\theta_{\rm SH}(iD_s q_1\mp v)q
%    e^{iqx}/\sqrt 2}{ D_{s}(q^{2}+q_{1}^{2})\mp ivq_1-D_aq^2}\nonumber\\
%  &&\mbox{}\times[e^{-D_aq^2 t}-e^{-D_{s}(q^{2}+q_{1}^{2})t\pm ivq_1t}].\label{Smp2}
%\end{eqnarray}

%\subsection{Numerical Results for General Cases}
%In this section, we show some numerical results for general values of $\alpha$ and $\beta$. Focusing on the spin-charge conversion, we use the initial condition as an optical-created electron-hole grating in $n$-type GaAs quantum well, which could be set up by transient grating technique, i.e., based on the interference between two incident laser beams polarized in the same direction.~\cite{Cameron96,Weber05,Yang2011a} 

\subsection{Spin-current swapping again} \label{SCSgradient}
We conclude this section by showing how the spin-current swapping effect, which vanishes  in the homogeneous situation of subsection~\ref{SC_swapping}, can be observed in an inhomogeneous situation, such as the one proposed by Lifshitz and D'yakonov in Ref.~\onlinecite{Lifshits09}.
At variance with subsection~\ref{SC_swapping} we assume that a spin current $J^x_x$ is injected into an {\it unpolarized} 2DEG in a (001) quantum well, coming from a ferromagnetic electrode polarized along the $x$ axis.  In the vicinity of the contact an inhomogeneous spin accumulation $S^x$ is induced, which decays on the scale of the spin-diffusion length (we neglect the ferromagnetic proximity effect).  The inhomogeneous spin accumulation drives an additional diffusion spin-current, which must be added to the drift spin-current considered in subsection~\ref{SC_swapping}.  
For simplicity, we ignore the intrinsic SOC, i.e., we set $\lambda_1=\lambda_2=0$.  However, we retain the SOC with the electric field $E_x$ (given by $A^z_y$ in Eq.~(\ref{QW001})) and, of course, the spin-current swapping term due to extrinsic impurities.   From  Eq.~(\ref{eqsc}), after expanding the covariant derivative and setting $v_y=0$ and $S^y=S^z=0$, we obtain
\ber
J_x^x &=& -(v_x+D\partial_x)S^x-\kappa J_y^y\nn\\
J_y^y &=& -2Dm\alpha'eE_x S^x-\kappa J_x^x\,.
\eer
Noting that $-2Dm\alpha'eE_x = -\kappa v_x$, we rewrite the second equation as
\be
J_y^y = -\kappa(v_xS^x+J_x^x)\,.
\ee
Solving the coupled equations for $J_x^x$ and $J_y^y$ yields
\ber
J_x^x &=& -v_xS^x-\frac{D}{1-\kappa^2} \partial_xS^x\nn\\
J_y^y &=& \frac{\kappa D}{1-\kappa^2} \partial_xS^x\,.
\eer
Thus, the $J_x^x$ component of the spin current remains, up to first order in $\kappa$ equal to the primary current injected by the ferromagnetic electrode.  But the spin-current swapping effect manifests itself in the appearance of a $J_y^y$ component of the spin current, which is proportional to the diffusion part of the primary spin current: $J_y^y \simeq \kappa D\partial_xS^x$.   This should be observable.

\section{Summary}
We have derived the microscopic spin kinetic equation in periodically modulated two dimensional electron liquids from non-equilibrium Green's function approach. We include the spin-orbit couplings due not only to the band structure, but also to the external electric field and the (non-magnetic) impurities.  Starting from the solution of the spin kinetic equation obtained from a perturbation expansion in the relaxation time approximation, we have derived a set of complete drift-diffusion equations for the charge and spin densities, in the presence of  an external electric field and a grating wave vector in arbitrary directions. We find that in the drift-diffusion equations the three mechanisms of  spin Hall effect, i.e., skew scattering, side-jump  and the intrinsic mechanism, can be combined together into a single spin Hall angle. Moreover, we also derive explicit expressions for the charge current and the spin current and, by combining them with the drift-diffusion equations, we analyze Edelstein effect, spin Hall effect and their inverses, as well as the spin current swapping effect. We then apply our theory to the study of  spin and density gratings in GaAs quantum wells. We recover the results of Doppler velocimetry experiments when the grating wave vector is parallel to the external electric field. For the grating wave vectors perpendicular to the electric field, we predict the conversion from electron-hole density grating to spin grating due to spin Hall effect. We show that single-spin-helical-mode pumping can be realized via spin Hall effect in (001) GaAs quantum well with identical Rashba and Dresselhaus coefficients.  We also show that the spin-current-swapping effect vanishes in a homogeneous situation, but can be detected in a spin injection experiment.

Missing from the analysis is the effect of electron-electron scattering on the spin conductivity and the spin diffusion constant (the so-called spin Coulomb drag~\cite{DAmico00,Flensberg01,Weber05}).  This can be included without difficulty: the connection between spin currents and electric fields in a spin-polarized interacting electron gas will be considered elsewhere.
Finally, we note that  it will be interesting to apply the present formalism to the theoretical analysis of surface acoustic wave experiments as done recently in Ref.\onlinecite{wanner14}.

%In summary, we have derived the kinetic equation from non-equilibrium Green's function by explicitly including the band SOC, external electric field and disorder, as well as the SOC due to the external field and disorder. Within Born approximation, all the relevant corrections, such as, energy correction due to the band SOC, spin precession around impurity and effective diffusion, are introduced. The combination effect of external electric field and disorder-induced SOC is also taking into account, which is essential to complete side-jump contribution. By employing the relaxation time approximation and perturbation expansion, we derived the density-spin coupled  diffusion matrix. We showed that the side-jump effect and skew scattering share the same structure in diffusion matrix, therefore, one can treat them together as a single extrinsic spin Hall velocity. However, the intrinsic mechanism differs from them by a factor in different matrix elements. We also discussed the spin grating produced from density grating in detail. 

\begin{acknowledgments}
We  acknowledge support  from NSF Grant No. DMR-1104788 (KS) and from the SFI Grant 08-IN.1-I1869 and the Istituto Italiano di Tecnologia under the SEED project grant No. 259 SIMBEDD (GV).    One of us (GV) thanks  the Donostia International Physics Center for hospitality and support during the completion of this work.  We especially thank Ilya Tokatly for many passionate and useful discussions about the fundamental structure of the SU(2) theory.  

% RR acknowledges partial support from EU through Grant. No. PITN-GA-2009-234970.
%We gratefully acknowledge support  from NSF Grant No. DMR-1104788.
%We thank Roberto Raimondi for valuable discussions.
\end{acknowledgments}
\appendix
\section{Matrices in Eq.~(\ref{KE2})} \label{APP_Matrix}
Including spin precession, diffusion and drift terms, ${\cal K}_{\mathbf k}$ reads
\begin{eqnarray}
&&{\cal K}_{{\bf k}}=\nonumber\\
&&\left(\begin{array}{cccc}
\Omega & i\lambda_{2}\tau{q}_{y} & i\lambda_{1}\tau{q}_{x} & im\alpha'(\qv\times\vv)_z\\
i\lambda_{2}\tau{q}_{y} & \Omega & -2B_{z}\tau & 2B_{y}\tau\\
i\lambda_{1}\tau{q}_{x} & 2B_{z}\tau & \Omega & -2B_{x}\tau\\
im\alpha'(\qv\times\vv)_z & -2B_{y}\tau & 2B_{x}\tau & \Omega
\end{array}\right),\nonumber\\
\end{eqnarray}
where $\Omega=-i\omega\tau+i\frac{\tau}{m}\mathbf k\cdot{\mathbf q}-e\mathbf E\cdot\nabla_{\mathbf k}$ with $\omega$ and ${\bf q}$ being the Fourier conjugate variables with respect to $t$ and ${\bf r}$. Note that the spin-spin coupling components actually show the precession effect due to the effective SOC field, which has components  $B_{x}=-k_{y}\lambda_{2}$, $B_{y}=-k_{x}\lambda_{1}$, $B_{z}=-e\alpha' (\kv\times\Ev)_z$.
%$B_{z}=-(k_{x}\gamma_{y}-k_{y}\gamma_{x})$.
% with $\gamma_i=\alpha'eE_i$. 

The second matrix, ${\cal T}_{\mathbf k}$, which describes the energy correction due to SOC, shows the density-spin coupling from the collision term $I_{\mathbf k}^{(a)}$ and $I_{\mathbf k}^{(c)}$ as
\begin{eqnarray}
{\cal T}_{{\bf k}} & = & \left(\begin{array}{cccc}
0 & -B_{x}\partial_{\epsilon_{k}} & -B_{y}\partial_{\epsilon_{k}} & -2B_{z}\partial_{\epsilon_{k}}\\
-B_{x}\partial_{\epsilon_{k}} & 0 & 0 & 0\\
-B_{y}\partial_{\epsilon_{k}} & 0 & 0 & 0\\
-2B_{z}\partial_{\epsilon_{k}} & 0 & 0 & 0
\end{array}\right).
\end{eqnarray}
Note that one half of the coupling between the density and the spin component along the $z$ direction comes from $I_{\mathbf k}^{(a)}$, while the other half comes from $I_{\mathbf k}^{(c)}$ as mentioned in the main text. 
%These two contributions together give the total side jump effect. %One can see that the relaxation time approximation technique makes both ${cal K}_{\bf k}$ and ${\cal T}_{\bf k}$ dependent only on ${\mathbf k}$. However, for the other terms, the relative charge both initial and final momentums are relevant . Specifically, we have

The matrix ${\cal M}_{\mathbf k,\mathbf k'}$, depending on the relative angle between the incoming momentum and the outgoing momentum in an electron-impurity scattering process, has three contributions. The first piece comes from the second term in $I^{(0)}_{\mathbf k}$ (see Eq.~(\ref{IK0})), corresponding to the spin current swapping term, which leads to
\begin{equation}
{\cal M}^{\rm sw}_{\mathbf{k},\mathbf{k}'}={\alpha'}\left(\begin{array}{cccc}
0 & 0 & 0 & 0\\
0 & 0 & -2(k_{x}k_{y}'-k_{y}k_{x}') & 0\\
0 & 2(k_{x}k_{y}'-k_{y}k_{x}') & 0 & 0\\
0 & 0 & 0 & 0
\end{array}\right)\,.
\end{equation}
The second piece, coming from the third term in $I^{(0)}_{\mathbf k}$, can be expressed as
\begin{equation}
{\cal M}^{\rm inh}_{\mathbf{k},\mathbf{k}'}={\alpha'}\left(\begin{array}{cccc}
0 & 0 & 0 & i[\mathbf{q}\times(\mathbf{k}-\mathbf{k}')]_{z}\\
0 & 0 & 0 & 0\\
0 & 0 & 0 & 0\\{}
i[\mathbf{q}\times(\mathbf{k}-\mathbf{k}')]_{z} & 0 & 0 & 0
\end{array}\right)
\end{equation}
where the superscript ``inh'' suggests that this term  reflects the effect of the spatial inhomogeneity during the scattering. The last contribution corresponds to the skew scattering $I_{\bf k}^{\rm ss}$ (Eq.~(\ref{ISS}))  resulting in 
\begin{equation}
{\cal M}_{\mathbf k,\mathbf k'}^{\rm ss}=\alpha_{ss}\left(\begin{array}{cccc}
    0 & 0 & 0 & -(k_x k_y'-k_yk_x')\\
    0 & 0 & 0 & 0\\
    0 & 0 & 0 & 0\\
    -(k_x k_y'-k_yk_x') & 0 & 0 &
0\end{array}\right).\end{equation}
with $\alpha_{ss}= \tau n_{i}\alpha' \left({mv_0}\right)^3(2\pi m)^{-1}$. Thus, the matrix ${\cal M}_{\mathbf k,\mathbf k'}$ in Eq.~(\ref{KE2}) is expressed by
\begin{equation}
  {\cal M}_{\mathbf k,\mathbf k'}={\cal M}^{\rm sw}_{\mathbf{k},\mathbf{k}'}+{\cal M}^{\rm inh}_{\mathbf{k},\mathbf{k}'}+{\cal M}_{\mathbf k,\mathbf k'}^{\rm ss}
\end{equation}
Notice that ${\cal M}_{\mathbf k,\mathbf k'}$ is proportional to $\alpha'$.

In our kinetic equation~(\ref{KE2}), $I^{(b)}_{\bf k}$ is not written in the form of a matrix, because, as explained in Sec.~\ref{SecKE}, the relaxation time approximation is inapplicable to it.
%\section{Drift-diffusion equations}
%From {\it SU(2)} theory, the drift-diffusion is given by
%\begin{eqnarray}
%  \partial_t S^a &=&-iq_i{J}_i^a+2\epsilon_{abc}A^b_i{J}^c_i-S^a/\tau_{sa}^{\rm EY},\nonumber\\
%  J_i^a&=&-(v_i+i D q_i)  S^a +2D\epsilon_{abc}A_i^bS^c - \theta_{\rm SH} \epsilon_{ijz} \zeta_z^aJ_j,\nonumber\\
%  J_i&=&-(v_i+iDq_i ) N - \theta_{\rm SH} \epsilon_{ijz}\zeta_z^aJ_j^a,
%\end{eqnarray}
%with $\zeta_z^z=1$ and $\zeta_z^{x(y)}=0$. The non-zero $SU(2)$ gauge field $A_x^y=m\lambda_1$ and $A_y^x=m\lambda_2$.
%%%%%%%%%%%%%%%%%%%%%%%%%%%%%%%%%%%%%%%%%%%%%%%%%%

\section{Derivation of the drift-diffusion equations for the densities} \label{APP_DDE}
In this section, we present the details of the derivation of the drift-diffusion equations for the densities.  Intuitively, the total diffusion matrix in Eq.~(\ref{Dq}) can be separated into two parts, the intrinsic part and the extrinsic one, according to the extrinsic SOC parameter $\alpha'$. That is
\be
{\cal D}={\cal D}^{\rm int}+{\cal D}^{\rm ext},
\ee
where the intrinsic contribution ${\cal D}^{\rm int}$ corresponds to the zero-th order term in $\alpha'$, equal to ${\cal D}_{\alpha'=0}$. The extrinsic part ${\cal D}^{\rm ext}$ in principle contains all high order contributions in $\alpha'$, however, because of the small value of $\alpha'$, it is sufficient to include only the first order term. 

Specifically, the intrinsic diffusion matrix is given by
\be D^{\rm int}=(1/\tau)\left[{\cal I}-\langle({\cal I}+{\cal K}_{\mathbf k})^{-1}({\cal I}+{\cal T}_{\mathbf k})\rangle\right]_{\omega=0,\alpha'=0},
\ee
%In the presence of an external electric field and  a grating modulation, there are three scale parameters with the units of a wave vector, i.e., $q_{x,y}=v_{x,y}/D$, and $2m\lambda_{1,2}$. 
%{\ROB I do not understand the equality: may be 
%$q_{x,y}$, $v_{x,y}/D$, and $2m\lambda_{1,2}$ so that the scales are three.}
%$1/q$ is the length scale over which the electric field changes the energy of an electron by a quantity comparable to the Fermi energy; $1/(m\lambda)$ is the length scale over which the spin of electron traveling at the Fermi velocity  completes a full round of precession in the spin-orbit field.   
%{\ROB There appears to be a repetition}
In the presence of an external electric field and  a grating modulation, there are three scale parameters with the units of a wave vector, i.e., $q_{x,y}$, $v_{x,y}/D$, and $2m\lambda_{1,2}$. $1/q$ is the grating wave length. $D/v$ is the length scale over which the electric field changes the energy of an electron by a quantity comparable to the Fermi energy; $1/(m\lambda)$ is the length scale over which the spin of electron traveling at the Fermi velocity  completes a full round of precession in the spin-orbit field.
In order to make our theory applicable for general values of the  ratios between these  scale parameters, we treat them on equal footing in the calculation and define a characteristic inverse length scale $l^{-1}$ as the largest of the three wave vectors.   We focus on the diffusive limit, i.e., the mean free path $v_F\tau\ll l$ and $E_F\tau\gg 1$, and we do a perturbation expansion with respect to ${\cal T}_{\mathbf k}$ and ${\cal K}_{\mathbf k}$. We find that the leading order of the density-density and spin-spin couplings is given by $l^{-2}$, while the first non-vanishing spin-density couplings are of the order of $l^{-4}$. Specifically, the spin-spin and density-density couplings are given by $\langle -{\cal K}_{\mathbf k}^2\rangle$ and the spin-density couplings are carried by  $\langle {\cal K}_{\mathbf k}^3+{\cal K}_{\mathbf k}^3{\cal T}_{\mathbf k}\rangle$ (the lowest order terms $\langle{\cal K}_{\mathbf k}\rangle$ and $\langle{\cal K}_{\mathbf k}{\cal T}_{\mathbf k}\rangle$ cancel against each other). Collecting all the relevant contributions in the leading order, we obtain the diffusion matrix due to the  intrinsic mechanism
\begin{widetext}
\begin{eqnarray}
%{\cal D}^{\rm int}&	\simeq &(1/\tau)	\langle{\cal K}+{\cal K}{\cal T}-{\cal K}^{2}-{\cal K}^{2}{\cal T}+{\cal K}^{3}+{\cal K}^{3}{\cal T}\rangle_{\omega=0,\alpha'=0}\nonumber \\
{\cal D}^{\rm int}	&=&\left(\begin{array}{cccc}
Dq^{2}-i\mathbf{q}\cdot\mathbf{v} & -i\theta^{\rm int}_{{\rm SH}}Dq_{y}q_1 & -i\theta^{\rm int}_{{\rm SH}}Dq_{x}q_2 & -i\theta^{\rm int}_{{\rm SH}}(\mathbf{v}\times\mathbf{q})_{z}\\
-\theta^{\rm int}_{{\rm SH}}(v_{y}+iDq_{y})q_1 & Dq^{2}-i\mathbf{q}\cdot\mathbf{v}+\frac{1}{\tau^{\rm DP}_{sx}} & 0 & (i2Dq_{x}+v_{x})q_1\\
-\theta^{\rm int}_{{\rm SH}}(v_{x}+iDq_{x})q_2 & 0 & Dq^{2}-i\mathbf{q}\cdot\mathbf{v}+\frac{1}{\tau^{\rm DP}_{sy}} & -(i2Dq_{y}+v_{y})q_2\\
-i\theta^{\rm int}_{{\rm SH}}(\mathbf{v}\times\mathbf{q})_{z} & -(i2Dq_{x}+v_{x})q_1 & (i2Dq_{y}+v_{y})q_2 & Dq^{2}-i\mathbf{q}\cdot\mathbf{v}+\frac{1}{\tau^{\rm DP}_{sz}}
\end{array}\right).
 \end{eqnarray}
The notation is defined in main text.

The extrinsic part of the diffusion matrix originates from two SOC sources, i.e., the one due to the external electric field and the other due to impurity potential. The contribution from the former one can be carried out from $(1/\tau)[{\cal I}-\langle({\cal I}+{\cal K}_{\mathbf k})^{-1}({\cal I}+{\cal T}_{\mathbf k})\rangle]_{\omega=0}-{\cal D}^{\rm  int}$, leading to
\begin{eqnarray}
{  \cal D}^{\rm ext,E}&\simeq &(1/\tau)[\langle{\cal K}_{\bf k}+{\cal K}_{\bf k}{\cal T}_{\bf k}-{\cal K}^{2}_{\bf k}\rangle_{\omega=0}-\langle{\cal K}_{\bf k}+{\cal K}_{\bf k}{\cal T}_{\bf k}-{\cal K}^{2}_{\bf k}\rangle_{\omega=0,\alpha'=0}]\nonumber\\
&=&\left(\begin{array}{cccc}
0 & 0 & 0 & -i(\theta^{\rm sj}_{{\rm SH}}/2)(\mathbf{v}\times\mathbf{q})_{z}\\
-(\theta^{\rm sj}_{{\rm SH}}/2)v_{y}q_1 & 0 & i2\kappa(\mathbf{v}\times\mathbf{q})_{z} & \kappa v_x q_2\\
-(\theta^{\rm sj}_{{\rm SH}}/2)v_{x}q_2 &-i2\kappa(\mathbf{v}\times\mathbf{q})_{z} & 0 & -\kappa v_y q_1\\
-i(\theta^{\rm sj}_{{\rm SH}}/2)(\mathbf{v}\times\mathbf{q})_{z} & \kappa v_x q_2 &-\kappa v_y q_1 & 0
\end{array}\right).
\end{eqnarray}
where the spin-spin coupling is from $\langle-{\cal K}_{\bf k}^2\rangle$.
Similarly, we can calculate the contribution from the extrinsic SOC due to impurity potential. By substituting the three $\cal M$ matrices into $-\frac{1}{\tau}\left\langle ({\cal I}+{\cal K}_{\bf k'})^{-1}{\cal   M}_{\mathbf k',\mathbf k}  ({\cal I}+{\cal K}_{\bf k})^{-1} ({\cal I+T}_{\bf
    k})\right\rangle_{\omega=0}$,  we obtain
\begin{eqnarray}
{\cal D}^{\rm ext, sw}&\simeq &(1/\tau)\langle{\cal K}_{\mathbf{k}'}{\cal M}^{\rm sw}_{\mathbf{k}',\mathbf{k}}{\cal T}_{\mathbf{k}}-{\cal K}_{\mathbf{k}'}{\cal M}^{\rm sw}_{\mathbf{k}',\mathbf{k}}{\cal K}_{\mathbf{k}}\rangle_{\omega=0}\nonumber \\
&=&\left(\begin{array}{cccc}
0 & 0 & 0 & 0\\
i\theta^{\rm sj}_{{\rm SH}}Dq_{y}q_1 & 0 & -i\kappa(\mathbf{v}\times\mathbf{q})_{z} & i\kappa Dq_x q_2\\
i\theta^{\rm sj}_{{\rm SH}}Dq_{x}q_2 &i\kappa(\mathbf{v}\times\mathbf{q})_{z} & 0 & -i\kappa Dq_y q_1\\
0 &i\kappa Dq_x q_2&-i\kappa Dq_y q_1 & 2\kappa Dq_1 q_2
\end{array}\right),
 \end{eqnarray}
\begin{eqnarray}
{\cal D}^{\rm ext,inh}&\simeq &(1/\tau)\langle{\cal K}_{\mathbf{k}'}{\cal M}^{\rm inh}_{\mathbf{k}',\mathbf{k}}-{\cal M}^{\rm inh}_{\mathbf{k}',\mathbf{k}}{\cal T}_{\mathbf{k}}+{\cal M}^{\rm inh}_{\mathbf{k}',\mathbf{k}}{\cal K}_{\mathbf{k}}\rangle_{\omega=0}\nonumber \\
&=&\left(\begin{array}{cccc}
0 & -i(\theta^{\rm sj}_{{\rm SH}}/2)Dq_{y}q_1 & -i(\theta^{\rm sj}_{{\rm SH}}/2)Dq_{x}q_2  & -i(\theta^{\rm sj}_{{\rm SH}}/2)(\mathbf{v}\times\mathbf{q})_{z}\\
-i(\theta^{\rm sj}_{{\rm SH}}/2)Dq_{y}q_1 & 0 & 0 & 0\\
-i(\theta^{\rm sj}_{{\rm SH}}/2)Dq_{x}q_2 &0 & 0 & 0\\
-i(\theta^{\rm sj}_{{\rm SH}}/2)(\mathbf{v}\times\mathbf{q})_{z} &0&0 & 0
\end{array}\right),
 \end{eqnarray}
\begin{eqnarray}
{\cal D}^{\rm ext,ss}&	\simeq &	-(1/\tau)\langle {\cal K}_{\mathbf k'}{\cal M}_{\mathbf k',\mathbf k}^{\rm ss} {\cal K}_{\mathbf k}\rangle_{\omega=0} \nonumber\\
	&=&\left(\begin{array}{cccc}
0 & -i\theta^{\rm ss}_{{\rm SH}}Dq_{y}q_1 & -i\theta^{\rm ss}_{{\rm SH}}Dq_{x}q_2 & -i\theta^{\rm ss}_{{\rm SH}}(\mathbf{v}\times\mathbf{q})_{z}\\
-\theta^{\rm ss}_{{\rm SH}}(v_{y}+iDq_{y})q_1 & 0 & 0 & 0\\
-\theta^{\rm ss}_{{\rm SH}}(v_{x}+iDq_{x})q_2 & 0 & 0 & 0\\
-i\theta^{\rm ss}_{{\rm SH}}(\mathbf{v}\times\mathbf{q})_{z} & 0 &0&0
\end{array}\right).
 \end{eqnarray}
To calculate the contribution from the spin-precession scattering term, i.e., the last term in Eq.~(\ref{Dq}), we substitute $\tilde g^i_{\mathbf k}\simeq (1+\tau e\mathbf E\cdot\nabla_{\mathbf k}-i\tau\mathbf q\cdot{\mathbf k}/{m}) g^i_{k}$ into $I^{(b)}$. After some straightforward calculation, we obtain the current from $\langle ({\cal K}_{\mathbf k}-{\cal I}) I_{\mathbf k}^{(b)}\rangle$ and rewrite the result in the form of diffusion matrix as
\begin{eqnarray}
{\cal D}^{{\rm ext},(b)}&=&\left(\begin{array}{cccc}
0 & -i(\theta^{\rm sj}_{{\rm SH}}/2)Dq_{y}q_1 & -i(\theta^{\rm sj}_{{\rm SH}}/2)Dq_{x}q_2 & 0\\
-(\theta^{\rm sj}_{{\rm SH}}/2)(3iDq_{y}+v_y)q_1 & 0 & 0 & 0\\
-(\theta^{\rm sj}_{{\rm SH}}/2)(3iDq_{x}+v_x)q_2 &0 & 0 & 0\\
0 &0&0 & 0
\end{array}\right).
\end{eqnarray}
Finally, we obtain the total extrinsic contribution in the diffusion matrix 
\be
{\cal D}^{\rm ext}={\cal D}^{\rm ext,E}+{\cal D}^{\rm ext,sw}+{\cal D}^{\rm ext,inh}+{\cal D}^{\rm ext,ss}+{\cal D}^{{\rm ext},(b)}
\ee
By collecting all the intrinsic and extrinsic pieces, we write out the final diffusion matrix shown in Eqs.~(\ref{DiffusionEquationMatrixForm}). Note that, the extrinsic contribution is discarded in Eqs.~(\ref{DiffusionEquationMatrixForm}) for the matrix element containing intrinsic contribution, by taking into account the fact the extrinsic SOC is weaker than the intrinsic one.

\section{Derivation of the drift-diffusion equations for the currents} \label{APP_current}
The goal of this appendix is to derive the transformation matrices $\hat J_x$ and $\hat J_y$, which connect the (spin) currents to the (spin) densities, so that one can obtain the currents directly from the densities via the equation
\be
{J}_i^j(\mathbf q) ={\hat J}_i^{jl}S^l_{\mathbf q}.
\ee
The transformation matrices, according to Eqs.~(\ref{gk_tot}) and (\ref{J_tot}), can be expressed as
\begin{equation}
\hat { J}_{x(y)}^{ij} = \left\langle {\cal J}_{x(y)\bf k} [{\cal I}+({\cal I}+{\cal K}_{\bf k'})^{-1}{\cal
      M}_{\mathbf k',\mathbf k}]  ({\cal I}+{\cal K}_{\bf k})^{-1} ({\cal I+T}_{\bf
    k})\right\rangle_{ij}
+\left.({\tau}/{S_{\mathbf q}^j})\left\langle {\cal J}_{x(y)\mathbf k}[({\cal I}+{\cal K}_{\bf k})^{-1}]^{il} I_{\mathbf{k}}^{(b),l}({\tilde {\bf g}}_{j\mathbf k},{\tilde {\bf g}}_{j\mathbf k^\prime})\right\rangle\right\vert_{\omega=0}.
\end{equation}
where the current matrices are given by
\be
{{\cal J}}^{ij}_{x(y)\mathbf k}=(1/4){\rm Tr}[\sigma^i\{\sigma^j,\tilde {v}_{x(y)\mathbf k}\}].
\ee
Specifically, we have
\begin{equation}
  {\cal J}_{x\mathbf k}=\left(\begin{array}{cccc}
    k_{x}/m & 0 & \lambda_{1} & \alpha'(eE_y+\frac{k_y}{\tau})\\
    0 & k_{x}/m & 0 & 0\\
    \lambda_{1} & 0 & k_{x}/m & 0\\
    \alpha'(eE_y+\frac{k_y}{\tau}) & 0 & 0 & k_{x}/m
  \end{array}\right),
\end{equation} and
\begin{equation}
  {\cal J}_{y\mathbf k}=\left(\begin{array}{cccc}
    k_{y}/m & \lambda_{2} & 0&-\alpha'(eE_x+\frac{k_x}{\tau})\\
    \lambda_{2} & k_{y}/m & 0 & 0\\
    0 & 0 & k_{y}/m & 0\\
    -\alpha'(eE_x+\frac{k_x}{\tau}) & 0 & 0 & k_{y}/m
  \end{array}\right),
\end{equation}
for the current flowing along the $x$ and $y$ directions, respectively. In the following, we take the current in the $x$ direction as an example to show the details of perturbation calculation.

By using the same technique introduced in Appendix~\ref{APP_DDE}, we obtain the intrinsic contribution
\begin{eqnarray}
  {\hat J}^{\rm int}_x&=&\langle {\cal J}_x({\cal I}+{\cal T}+{\cal K}^{2}+{\cal K}^{2}{\cal T}-{\cal K}_{\mathbf{}}-{\cal K}_{\mathbf{}}{\cal T}_{\mathbf{}})\rangle_{\omega=0,\alpha'=0}\nonumber\\
  &=&\left(\begin{array}{cccc}
-(iDq_{x}+v_{x}) & 0 & -\theta^{\rm int}_{{\rm SH}}Dq_2 & \theta^{\rm int}_{{\rm SH}} (iDq_{y}+v_{y})\\
0 & -(iDq_{x}+v_{x}) & 0 & Dq_1\\
0 & 0  & -(iDq_{x}+v_{x}) & 0\\
\theta^{\rm int}_{{\rm SH}} (iDq_{y}+v_{y}) & -Dq_1 & 0 & -(iDq_{x}+v_{x})
\end{array}\right).
\end{eqnarray}
\end{widetext}
With the same notation as used in the diffusion matrix, the relevant extrinsic terms are
${\hat J}_x^{\rm ext,E}=\langle {\cal J}_{x\mathbf k}({\cal I}+{\cal T}_{\mathbf k}-{\cal K}_{\mathbf k})\rangle_{\omega=0}-\langle {\cal J}_{x\mathbf k}({\cal I}+{\cal T}_{\mathbf k}-{\cal K}_{\mathbf k})\rangle_{\omega=0,\alpha'=0}$,
${\hat J}_x^{\rm ext,sw}=\langle{\cal J}_{x\mathbf k'}{\cal M}^{\rm sw}_{\mathbf{k}',\mathbf{k}}({\cal T}_{\mathbf{k}}-{\cal K}_{\mathbf{k}})\rangle_{\omega=0}$, 
${\hat J}_x^{\rm ext,inh}=\langle {\cal J}_{x\mathbf k'}{\cal M}_{\mathbf{k}',\mathbf{k}}^{\rm inh}\rangle $, and ${\hat J}_x^{\rm ext,ss}=-\langle{\cal J}_{x\mathbf k'}{\cal M}_{\mathbf{k}',\mathbf{k}}{\cal K}_{\mathbf{k}}\rangle_{\omega=0}$, resulting in
\be
  {\hat J}_{x}^{\rm ext,E}=\left(\begin{array}{cccc}
0 & 0 & 0 & ({\theta_{\rm SH}^{\rm sj}}/{2})v_{y}\\
0 & 0 & -\kappa v_{y} & 0\\
0 & \kappa v_{y} & 0 & 0\\
({\theta_{\rm SH}^{\rm sj}}/{2})v_{y} & 0 & 0 & 0
\end{array}\right),
\ee
\be
  {\hat J}_{x}^{\rm ext,sw}=\left(\begin{array}{cccc}
0 & 0 & 0 & 0\\
0 & 0 & \kappa(iDq_{y}+v_{y}) & \kappa Dq_2\\
{\theta_{\rm SH}^{\rm sj}}Dq_2 & -\kappa(iDq_{y}+v_{y}) & 0 & 0\\
0 & 0 & 0 & 0
\end{array}\right),
\ee
\be
  {\hat J}_{x}^{\rm ext,inh}=\frac{\theta_{\rm SH}^{\rm sj}}{2}\left(\begin{array}{cccc}
0 & 0 & 0 & iDq_{y}\\
0 & 0 & 0 & 0\\
0 & 0 & 0 & 0\\
iDq_{y} & 0 & 0 & 0
\end{array}\right),
\ee
\be
{\hat J}^{\rm ext, ss}_x=\left(\begin{array}{cccc}
0 & 0 & -\theta^{\rm ss}_{{\rm SH}}Dq_2 & \theta^{\rm ss}_{{\rm SH}} (iDq_{y}+v_{y})\\
0 & 0 & 0 & 0\\
0 & 0  & 0 & 0\\
\theta^{\rm ss}_{{\rm SH}} (iDq_{y}+v_{y}) & 0 & 0 & 0
\end{array}\right).
\ee
The (spin) current induced by the spin-precession scattering term, at the leading order, can be directly calculated from $({J}_{x}^{{\rm ext},(b)})^j=\langle(k_x/m)\tau I^{(b),j}\rangle$ by substituting $\tilde g^i_{\mathbf k}\simeq g^i_{k}$. The result in the form of matrix leads to
\begin{equation}
  {\hat J}_{x}^{{\rm ext}, (b)}={\theta_{\rm SH}^{\rm sj}}
\left(\begin{array}{cccc}
0 & 0 & -Dq_2 & 0\\
0 & 0 & 0 & 0\\
-Dq_2 & 0 & 0 & 0\\
0 & 0 & 0 & 0
\end{array}\right).
\end{equation}
Then the extrinsic mechanisms totally contribute to the transformation matrices
\be
   \hat{J}_x^{\rm ext}={\hat J}_x^{\rm ext,E}+{\hat J}_x^{\rm ext,sw}+{\hat J}_x^{\rm ext,inh}+{\hat J}_x^{\rm ext,ss}+{\hat J}_x^{{\rm ext},(b)}.
\ee
Note that in the final result in Eqs.~(\ref{J_x}) and (\ref{J_y}), only the leading term in each matrix element is retained.
%By adding the above four terms up, we find the spin-charge coupling due to side-jump effect in the current matrices is also present in the same structure as skew scattering and intrinsic mechanism. 
%However, the prefactor, instead of the total spin Hall angle due to side-jump effect, is found to be $\theta_{\rm SH}^{\rm sj}/2$, which means that only half of the side-jump effect is included. As explained in the main text, the other half can be sought from the anomalous velocity due to impurity-induced SOC.~\cite{Raimondi_AnnPhys12} This part can be calculated from ${{\hat J}_{x,\rm an}^{\rm sj}}=\langle{\cal J}'_{x\mathbf k}(-{\cal K}_{\mathbf k})\rangle$ with the current matrix of anomalous velocity due to impurity-induced SOC is given by
%\begin{equation}
%  {\cal J}'_{x\mathbf k}=\left(\begin{array}{cccc}
%0 & 0 & 0 & {\alpha'}k_{y}/\tau\\
%0 & 0 & 0 & 0\\
%0 & 0 & 0 & 0\\
%{\alpha'}k_{y}/\tau & 0 & 0 & 0
%\end{array}\right)
%\end{equation}

\section{Derivation of spin-current swapping} \label{APP_SCS}
As mentioned in the main text, the spin current swapping in our theory is included as the second term in the collision integral $I_{\mathbf k}^{(0)}$, i.e., $-i\alpha'({2\pi}{\tau})^{-1}\int {d\theta_{\mathbf k'}}[{\bg\sigma}\cdot\mathbf{k}\times\mathbf{k}^{\prime},\rho_{\mathbf{k}^{\prime}}]$, and appears already in the first Born approximation.  In the leading order, the correction in the steady-state density matrix due to spin current swapping term is given by
\be
\delta g_{\bf k}^{{\rm sw},j}=(2\alpha'/N_0)\sum_{lmn}\epsilon^{zlj}\epsilon^{zmn}k_m [J_n^l]^{(0)}.
\ee
with the primary spin current $[J_n^l]^{(0)}=\sum_{\mathbf k'} k^\prime_n g_{\bf k'}^l$. One then can calculate the swapped spin current from
\begin{eqnarray}
[J_i^j]^{\rm SCS}&\simeq& \sum_{\mathbf k} (k_i/m) g_{\bf k}^{{\rm sw},j}\nonumber\\
&=& \alpha'k_F^2\sum_{ln}\epsilon^{zlj}\epsilon^{zin}[J_n^l]^{(0)}\nonumber\\
&=&\alpha'k_F^2\left([J_j^i]^{(0)}-\delta_{ij}\sum_l [J_l^l]^{(0)}\right),
\end{eqnarray}
whose symmetry is consistent with previous work.~\cite{Lifshits09} Here, the coefficient of spin current swapping reads $\kappa=\alpha' k_F^2$.

\bibliography{SJ.bib}
\bibliographystyle{prsty}
\end{document}